\documentclass{cimento}

\usepackage{graphicx}  
\usepackage[pagewise]{lineno}

\title{Dissipative many-body physics of cold Rydberg atoms}
\author{O.~Morsch\from{ins:x}\ETC,
I.~Lesanovsky\from{ins:y}\from{ins:z},
\thanks{Any footnote to author.}}
\instlist{\inst{ins:x} INO-CNR and Dipartimento di Fisica, Largo Pontecorvo 3, 56127 Pisa, Italy
  \inst{ins:y}School of Physics and Astronomy, University of Nottingham, Nottingham, NG7 2RD, United Kingdom
  \inst{ins:z}Centre for the Mathematics and Theoretical Physics of Quantum Non-equilibrium Systems,
University of Nottingham, Nottingham NG7 2RD, United Kingdom}

\begin{document}

\maketitle

\begin{abstract}
In the last twenty years, Rydberg atoms have become a versatile and much studied system for implementing quantum many-body systems in the framework of quantum computation and quantum simulation. However, even in the absence of coherent evolution Rydberg systems exhibit interesting and non-trivial many-body phenomena such as kinetic constraints and non-equilibrium phase transitions that are relevant in a number of research fields. Here we review our recent work on such systems, where dissipation leads to incoherent dynamics and also to population decay. We show that those two effects, together with the strong interactions between Rydberg atoms, give rise to a number of intriguing phenomena that make cold Rydberg atoms an attractive test-bed for classical many-body processes and quantum generalizations thereof.  
\end{abstract}

\section{Introduction} Atoms excited to high-lying energy states (with principal quantum number larger than $\approx 15$) are commonly known as Rydberg atoms \cite{Gallagher_2005}. Compared to atoms in the ground or low-lying excited states, they have considerably longer lifetimes (on the order of hundreds of microseconds as opposed to nanoseconds) and larger electric polarizability. The latter property leads to small critical electric fields for field ionization as well as large van der Waals and dipole-dipole interactions between such atoms (several orders of magnitude larger than those of ground state atoms). Rydberg atoms have been studied for several decades, with a particularly productive period in the 1970's just after the invention of the laser \cite{Cooke_1980,Raimond_1981}. More recently, they have enjoyed another renaissance due to laser cooling, which made more accurate studies possible, and also due to the advent of quantum computation and quantum simulation, for which Rydberg atoms are a promising building block \cite{saffman_2010}. In fact, in the last two decades the peculiar properties of Rydberg atoms have been exploited for realizing two-qubit quantum gates \cite{Isenhower_2010} and for implementing quantum simulations of Ising systems \cite{Barredo_2016, Bernien_2017}. There, the strong and widely tunable long-range interactions between Rydberg atoms \cite{Comparat_2010} fill a gap in other quantum information approaches based on ultra-cold (neutral) atoms in the ground state \cite{Simon_2011,Bloch_2012}, which only interact through the much weaker contact interaction. The combination of controllability, strong interactions and long coherence times make Rydberg atoms promising candidates for the realization of future quantum information technologies. 

In spite of this justified focus on Rydberg systems exhibiting and exploiting quantum coherence, cold Rydberg atoms are not just interesting from the point of view of quantum many-body physics, but also offer valuable insight into a number of classical many-body phenomena. This is what we aim to show in this review of our recent experimental and theoretical work in Pisa and Nottingham. The systems we study are "classical" in the sense that dissipation introduced by coupling to the environment leads to decay of the quantum coherences and also to decay of the populations of the Rydberg states back to the ground state. In the regime in which decoherence is important but spontaneous decay is not yet relevant, the excitation dynamics to Rydberg states of a cloud of cold atoms can be viewed in terms of so-called kinetic constraints, which arise naturally from the interplay between Rydberg-Rydberg interactions and the detuning from resonance of the excitation laser. On the other hand, once the timescales become long enough for spontaneous decay to play a role, there is a competition between kinetically constrained excitation events and decay. In that case, critical phenomena related to absorbing-state phase transitions govern the properties of the system. 

Studying classical many-body phenonema using experimental and theoretical methods originating from the world of ultra-cold atoms and quantum optics may, at first sight, appear to be a less ambitious aim than realizing quantum many-body systems. Here we aim to show that, maybe somewhat surprisingly, studies of dissipative Rydberg systems can, indeed, yield valuable information on processes and phenomena typically associated with, e.g, soft matter such as glass formers \cite{Biroli_2013}, or even farther afield, such as wildfires and the spreading of infectious diseases \cite{Grassberger_1983}. Moreover, Rydberg systems offer the intriguing possibility to move away from this classical limit and to probe quantum generalisations of classical processes \cite{Perez-Espigares_2017} for which one may anticipate the emergence of new types of phases and transitions.

Taking the example of kinetic constraints, those are related to the dramatic slowdown often associated with the complex collective relaxation in classical many-body systems \cite{Lesanovsky_2013}. Essentially, kinetic constraints put a condition on the rate for a local transition to happen ({\em e.g.}, a particle inside a glass moving to a neighbouring position) that depends strongly on the local environment. This leads to strongly correlated collective and spatially inhomogeneous relaxation dynamics with properties that go beyond those of the stationary state. In spite of the simplicity of this concept, in practice it is not easy to establish a clear link between the microscopic processes inside a real material and the resulting (emerging) kinetic constraints. Conversely, idealized models involving explicit kinetic constraints are typically difficult to realize in an actual physical system. In that sense, it turns out that cold Rydberg atoms are an ideal testbed for such models as it is possible to clearly identify the microscopic processes and to implement them in a clean way in the Rydberg system.

Adding spontaneous decay as a competing process, cold Rydberg gases can be used to study absorbing state phase transitions occurring in some of the simplest models displaying critical behaviour  \cite{Hinrichsen_2000,Hinrichsen_2006}. Although conceptually simple, such models are of widespread interest and are the subject of current research across several disciplines. As in the case of kinetic constraints, clean implementations of the theoretical models are surprisingly difficult to achieve, and cold Rydberg atoms are a versatile platform for such models. 

The review is organized as follows. First, we describe our experimental setup and theoretical treatment of the Rydberg excitation dynamics in section 2. In section 3 we introduce the theoretical concept of kinetic constraints and show experimental results confirming their occurrence in a cold Rydberg gas. In particular, we demonstrate the fundamental difference between the blockade constraint (for resonant excitation) and the facilitation contstraint (for off-resonant excitation). Thereafter, we examine what happens when spontaneous decay is added to the system and provide evidence for an absorbing state phase transition. In section 4 we re-introduce quantum coherence into the theoretical description and provide ideas as  to how the phase transition may be affected by the coherences. We also discuss how those effects could be observed experimentally. Finally, in section 5 we summarize our results and give an outlook on future challenges and opportunities in this field. 

\section{Theoretical and experimental methods}
We realize a driven-dissipative many-body system using clouds of cold rubidium atoms and model our experiments using a minimal model of many-body effects in such a system. In this section we will briefly describe the experimental and theoretical methods necessary to understand the main part of this review.

Our experiments are performed with $^{87}\mathrm{Rb}$ atoms in a magneto-optical trap (MOT) containing up to $10^6$ atoms in roughly spherical clouds measuring between $50$ and $150\,\mathrm{\mu m}$ at temperature $T\approx 120\,\mathrm{\mu K}$ (measured using a release-and-recapture method). That temperature corresponds to a mean thermal velocity of the atoms of around $0.11\,\mathrm{\mu m/\mu s}$. As a consequence, while single excitation events to Rydberg states (details see below), occurring on a timescale of a few microseconds, can be considered to take place within the frozen gas regime ({\em i.e.}, the atoms can be considered stationary on the relevant length scales of the system), that approximation is no longer valid on the timescales of tens to hundreds of microseconds in the experiments on the dissipative phase transitions (sec. 3.4). 

\begin{figure}
\includegraphics[width=8 cm]{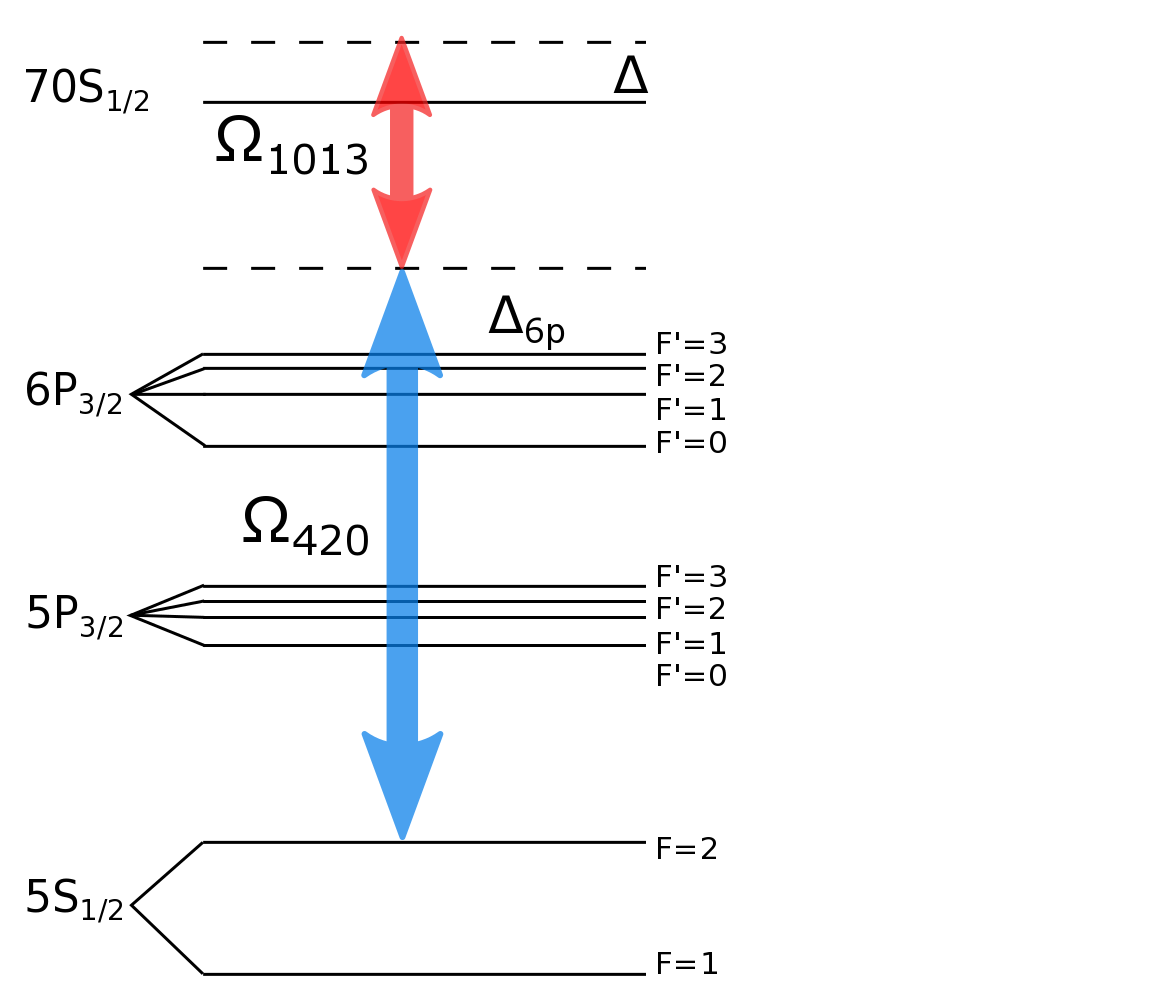}     
\caption{Energy levels of  $^{87}\mathrm{Rb}$. Rydberg states are excited using two laser beams via an intermediate state $6P$ that is detuned sufficiently from resonance. The detuning $\Delta$ from the Rydberg state is used to control the kinetic constraints described in sec. 3.  A laser beam resonant with the transition  $|5 S_{1/2}, F=2\rangle \rightarrow |5 P_{3/2}, F'=2\rangle$ can be used to depump atoms into the $|5 S_{1/2}, F=1\rangle $ that is not coupled to the Rydberg excitation lasers, thus creating lower effective densitites. From ref. \cite{Simonelliphd_2018}}\label{exp_levels}
\end{figure}

Based on the number of atoms in the MOT and its size, the atom densities in our experiments are typically on the order of a few $10^{10}\,\mathrm{cm}^{-3}$ to  $10^{11}\,\mathrm{cm}^{-3}$. In order to obtain smaller effective densities, which allow us to explore a greater range of interatomic distances, we use a laser pulse of around $2\,\mathrm{\mu s}$ duration resonant with the transition $|5 S_{1/2}, F=2\rangle \rightarrow |5 P_{3/2}, F'=2\rangle$, with the MOT repumping laser switched off (see fig. \ref{exp_levels}), that pumps a fraction of the atoms into the $|5 S_{1/2}, F=1\rangle $ hyperfine sublevel of the ground state~\cite{Valado_2015}. As our Rydberg excitation scheme resonantly couples only the  $|5 S_{1/2}, F=2\rangle $  sublevel to the Rydberg state, the fraction of atoms pumped to the $F=1$ sublevel (which lies $6.8\,\mathrm{GHz}$ below the $F=2$ level) does not participate in the excitation dynamics. In this way, the effective atom density can be reduced by up to a factor of $10^4$.

Rydberg states are excited (with the MOT beams switched off) using a two-photon scheme with a $420\,\mathrm{nm}$ laser (power up to $12\,\mathrm{mW}$, beam size between $7\,\mathrm{\mu m}$ and  $40\,\mathrm{\mu m}$),  blue-detuned by typically $\Delta_{6P}= 2\pi\times 0.5-1\,\mathrm{GHz}$ from the $|5 S_{1/2}, F=2\rangle\rightarrow |6 P_{3/2}, F'=3\rangle$ transition, and a $1013\,\mathrm{nm}$ laser (power up to $70\,\mathrm{mW}$, beam size $40\,\mathrm{\mu m}$)  providing the second photon for coupling to the $70 S$ state with a detuning $\Delta$. Both lasers have intrinsic linewidths around $2\pi\times 400\,\mathrm{kHz}$. The Rabi frequency for the two photon transition is
 \begin{equation}
\Omega= \sqrt{\frac{\Omega^2_{420} \Omega^2_{1013}}{4 \Delta^2_{6P}}},
 \end{equation}
 where $\Omega_{420}$ and $\Omega_{1013}$ are the Rabi frequencies of the single transitions. The individual Rabi frequencies were calibrated as follows. To determine $\Omega_{420}$, we measured the Autler-Townes splitting \cite{Viteau_2011} (using the second step laser as a probe), which yielded a maximum value of  $\Omega_{420}\approx 2\pi \times 40 \,\mathrm{MHz}$. The second step Rabi frequency was measured by resonantly de-exciting Rydberg atoms to the $6P$ intermediate state (see sec. 3.4). The frequency of the resulting (damped) Rabi oscillations was measured, giving a maximum value  $\Omega_{1013}\approx 2\pi \times 4 \,\mathrm{MHz}$. From these values, the maximum resonant two-photon Rabi frequency is found to be around $2\pi\times 250 \,\mathrm{kHz}$ for typical parameters of our experiment.

The two excitation lasers, together with the atomic density distribution in the MOT, define the effective interaction volume and geometry governing the excitation dynamics. In particular, we use two different sizes for the laser at $420\,\mathrm{nm}$, resulting in a three-dimensional interaction volume for a beam size of $40\,\mathrm{\mu m}$ and an effective one-dimensional geometry for a beam size  of $7\,\mathrm{\mu m}$. In the latter case, the fact that the radial size of the $420\,\mathrm{nm}$-beam is smaller than or comparable to the length scales that govern the many-body correlated dynamics of the system means that the creation of more than one excitation in the radial direction is strongly suppressed.

To theoretically describe the excitation dynamics, we focus here on the simplest possible scenario in which atoms are described within a two level (or effective spin-$1/2$) approximation. The two states we are considering are an electronic ground state $\left|g\right>=|5 S_{1/2}, F=2\rangle \equiv\left|\downarrow\right>$ and a high-lying Rydberg $S$-state denoted as $\left|r\right>\equiv\left|\uparrow\right>$.

Dynamics is driven by a laser field which couples the ground state $\left|\downarrow\right>$ to the excited state $\left|\uparrow\right>$. The coupling is parameterised by the Rabi frequency $\Omega$ (denoting the coupling strength) and the detuning $\Delta$ (characterising the frequency mismatch between the laser and the atomic transition). In our convention $\Delta>0$ when the laser is "blue-detuned". Employing the rotating wave approximation and introducing the Pauli spin operators $\sigma_k^\beta$ ($\beta=x,y,z$), as well as the projection operator on the Rydberg state, $n_k=(1+\sigma_k^z)/2$, we can now formulate the Hamiltonian of $N$ interacting atoms in the presence of the laser field:
\begin{eqnarray}
  H=\frac{\Omega}{2} \sum_{k=1}^{N} \sigma_x^k+\Delta \sum_{k=1}^{N} n_k+\frac{1}{2}\sum_{k,m=1}^{N} V_{km} n_k n_m. \label{eq:hamiltonian}
\end{eqnarray}
Here
\begin{eqnarray}
  V_{km} &=& \frac{C_6}{|\mathbf{r}_k-\mathbf{r}_m|^6}
\end{eqnarray}
parameterises the interaction strength between atoms at positions $\mathbf{r}_k$ and $\mathbf{r}_m$, with $C_6$ being the so-called dispersion coefficient related to the van der Waals interaction. For the $70S$ state of rubidium used in our experiments, $C_6=869.7 \,\mathrm{GHz}\,\mathrm{\mu m}^6$. The interaction is only non-zero provided that both the $k$-th and the $m$-th atom are simultaneously excited to the Rydberg states, {\em i.e.}, the expectation value of the number operator $n_k$ is non-zero for both atoms.

The formulation of the Rydberg problem in terms of Hamiltonian (\ref{eq:hamiltonian}) implies a decoupling of the internal dynamics from the external degrees of freedom. Often this is approximately true due to a separation of timescales provided by the frozen gas limit introduced above, and we will make that assumption in this section. In addition to the coherent dynamics effectuated by Hamiltonian (\ref{eq:hamiltonian}) there are dissipative processes which render the system open.  An established way of modelling these effects within the two-level approximation is to describe the dynamics of the density matrix $\rho$ of the Rydberg gas through a Markovian master equation of the form
\begin{eqnarray}
  \frac{\partial \rho}{\partial t}&=& -i \left[H,\rho\right]+\kappa\sum_k\left(\sigma_k^-\rho \sigma_k^+-\frac{1}{2}\left\{\sigma^+_k \sigma^-_k,\rho \right\}\right)+2\gamma\sum_k\left(n_k\rho n_k-\frac{1}{2}\left\{n_k,\rho \right\}\right).\label{eq:master_equation}
\end{eqnarray}
Here the first commutator term describes the coherent von-Neumann evolution of the density matrix. The terms proportional to $\kappa$ represent the (radiative) decay of the Rydberg state to the ground state (with the operators $\sigma_k^{\pm}=\frac{1}{2}[\sigma_k^x \pm i\sigma_k^y]$) and the terms proportional to $\gamma$ describe the dephasing of quantum superpositions between the two considered atomic states (note that the factor $2$ in front of $\gamma$ in eq. (\ref{eq:master_equation}) is simply chosen for convenience, so that $\gamma$ directly corresponds to the dephasing rate.).

This model description is an idealisation. This assumes that, indeed, radiative decay of the Rydberg state leads to an immediate relaxation to the electronic ground states. While this is typically the predominant channel, it ixxs known that there can be a cascaded decay which leads to the transient population of other Rydberg states. Furthermore, the model assumes that dephasing takes place for all atoms independently, {\em i.e.}, the dephasing noise is uncorrelated. Typically dephasing is a result of external field (or laser field) fluctuations that result in random atomic levels shifts. The spatial variations of these fields can occur on length scales that are larger than the typical interatomic distance, which would lead to correlated noise. In this sense the uncorrelated noise model employed here can only be regarded as a simple approximation, which, however, so far has turned out to usually yield a rather accurate description of experimental data \cite{Pfau_2012}.

\begin{figure}
\includegraphics[width=10 cm]{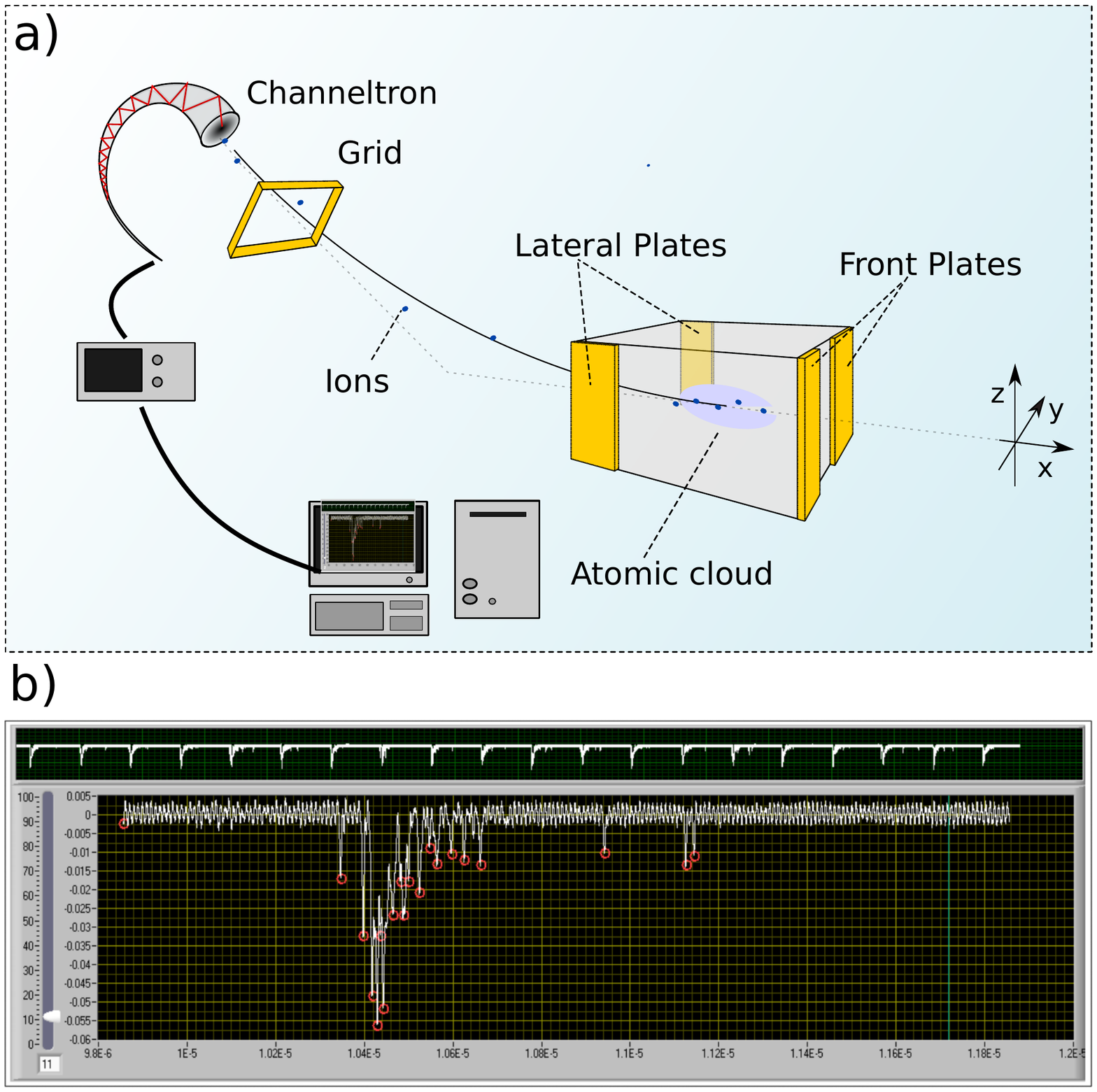}
\caption{Setup of the Rydberg experiment. In a) the field ionization plates and the channeltron used for the ionization and detection of the Rydberg atoms are shown. The channeltron signal (an example is shown in b)) is acqiured by an oscilloscope and analyzed on a computer. Each of the downward spikes in b) corresponds to the arrival of a single ion at the channeltron (from ref. \cite{Simonelliphd_2018}).}\label{exp_setup}
\end{figure}

Experimentally, once the excitation dynamics has taken place, we interrogate our system by applying a brief (on the order of a few microseconds) electric field pulse (see fig. \ref{exp_setup}) that field ionizes the Rydberg atoms (the critical field for the $70S$ state is around $20\,\mathrm{V/cm}$) and accelerates the resulting ions to a channeltron charge multiplier, where they are detected with an overall detection efficiency $\eta\approx 0.4$ (the experimental values reported in this review have been corrected for that detection efficiency) \cite{Viteau_2011}. The channeltron signal is acquired using a fast oscilloscope, and the number of detected ions in each experimental shot is determined using a peak-finding routine on a computer. Typical experimental runs consist of several hundred shots, the results of which are used to calculate the mean value and standard deviation of the number of detected ions, as well as histograms of the full counting statistics \cite{Malossi_2014}. In particular, from the mean $\langle N \rangle$ and standard deviation $\langle N^2 \rangle$, the Mandel $Q$-parameter
 \begin{equation}
Q=\frac{\langle N^2 \rangle}{\langle N \rangle^2} -1
 \end{equation}
is calculated, which yields information on the super- or sub-Poissonian character of the counting statistics (see sec. 3.2 for a detailed discussion).

\section{The incoherent driving regime: semi-classical dynamics}

\subsection{Kinetic constraints: Theory}
Although there is a fairly good understanding of the microscopic dynamics of Rydberg gases, through Eq. (\ref{eq:master_equation}), the theoretical exploration of collective behavior remains a challenge. This is owed to the lack of numerical methods permitting the study of large ensembles of quantum particles. A significant simplification of the problem - computationally and conceptually - can be achieved by focussing on the dissipative limit. On the one hand this allows to obtain effective equations of motion that are computationally tractable, even for large ensembles and in arbitrary dimensions. On the other hand this procedure very naturally leads to the notion of so-called kinetic constraints, {\em i.e.}, one manifestly observes that certain relaxation pathways are strongly suppressed, which in turn leads to a highly correlated and complex dynamical behavior.

In the strongly dissipative limit the dephasing rate $\gamma$ is the dominant energy scale. In this regime superposition states of atoms dephase rapidly (on a timescale $1/\gamma$) to become mixed states, {\em e.g.},
\begin{eqnarray}
  \frac{1}{2}(\mid\downarrow\rangle + \mid\uparrow\rangle)(\langle\downarrow\mid + \langle\uparrow\mid)\rightarrow  \frac{1}{2}\left[\mid\downarrow\rangle\langle\downarrow\mid + \mid\uparrow\rangle\langle\uparrow\mid\right].
\end{eqnarray}
Over sufficiently long timescales one thus no longer needs to consider coherences between atomic basis states. The dynamics is then described by a classical Master equation acting on the probability vector $\mathbf{p}$ which contains the populations of the classical atomic many-body basis states, {\em e.g.}, $\mid\downarrow\downarrow\downarrow...\rangle$, $\mid\uparrow\downarrow\downarrow...\rangle$, etc. This classical Master equation can be obtained via second order perturbation theory, which is discussed in detail in Refs. \cite{Cai13,Marcuzzi14,Degenfeld14}. It reads
\begin{eqnarray}
  \frac{\partial}{\partial t} \mathbf{p}=\sum_k \Gamma_k\left[\sigma^+_k-(1-n_k)\right]\mathbf{p}+\sum_k \Gamma_k\left[\sigma^-_k-n_k\right]\mathbf{p}+\kappa\left[\sigma^-_k-n_k\right]\mathbf{p},
\end{eqnarray}
with
\begin{eqnarray}
  \Gamma_k=\frac{\Omega^2}{2\gamma} \frac{1}{1+\left[\frac{\Delta}{\gamma}+\frac{C_6}{\gamma}\sum_{q\neq k} \frac{n_q}{|\mathrm{r}_k-\mathrm{r}_q|^6} \right]^2}.\label{eq:classical_master_equation}
\end{eqnarray}

Let us now consider for the sake of simplicity the case in which radiative decay is absent, {\em i.e.}, $\kappa=0$. In this case the dynamics of Eq. (\ref{eq:classical_master_equation}) is solely determined through kinetic constraints that are realized via the operator valued rates $\Gamma_k$ (due to their dependence on $n_q$). It is interesting to note that the stationary state $\mathbf{p}_\mathrm{ss}$ is in fact trivial. It is given by the product state
\begin{eqnarray}
  \mathbf{p}_\mathrm{ss}=\prod_k \frac{1}{2}\left(
                                              \begin{array}{c}
                                                1 \\
                                                1 \\
                                              \end{array}
                                            \right)
  \equiv\prod_k \frac{1}{2}\left[\mid\downarrow\rangle_k\langle\downarrow\mid + \mid\uparrow\rangle_k\langle\uparrow\mid\right] ,
\end{eqnarray}
in which each classical many-body configuration occurs with equal probability. The interesting aspect is, however, that the relaxation dynamics towards this stationary state is rather non-trivial. This fact makes kinetic constraints a relevant tool for the construction of models of glass formers \cite{Biroli_2013}. These substances can be also thought of as possessing a trivial stationary state, which is however never reached on accessible timescale due to the intricacy and slowness of the constrained relaxation dynamics.

How kinetic constraints work in Rydberg gases is probably best illustrated by considering two atoms: an excited one placed at the origin of the coordinate system ($\mathbf{r}_1=0$) and another other one in its ground state positioned at $\mathbf{r}_2=\mathbf{r}$. In this setting the rate for a state change of the second atom is given by
\begin{eqnarray}
  \Gamma_2=\frac{\Omega^2}{2\gamma} \frac{1}{1+R^{12}\left[\frac{1}{r_\mathrm{fac}^6}-\frac{1}{|\mathbf{r}|^6}\right]^2},
\end{eqnarray}
where we have introduced the dissipative blockade radius $R^6=\frac{C_6}{\gamma}$ and the facilitation radius --- which will be discussed later in detail --- defined through $r^6_\mathrm{fac}=-\frac{\gamma R^6}{\Delta}$.

\begin{figure}
\includegraphics[width=10 cm]{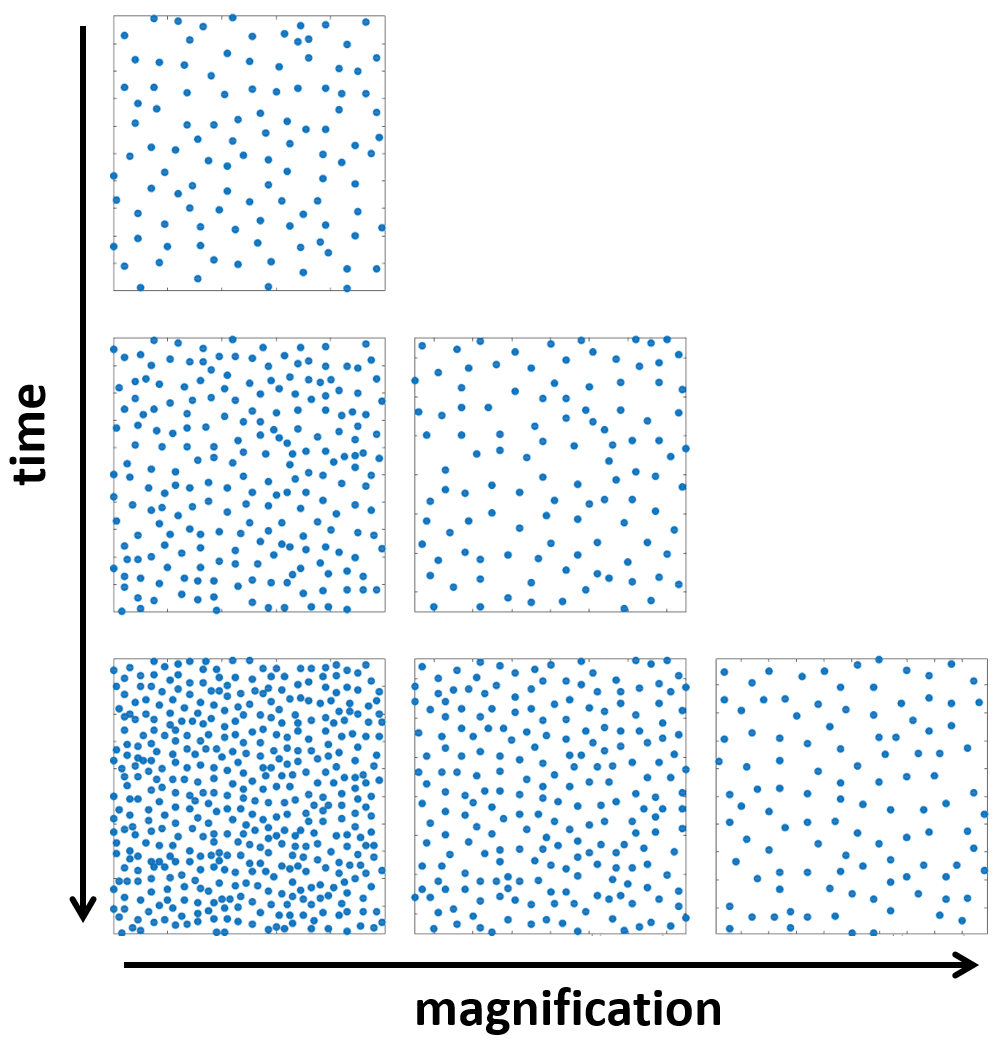}     
\caption{Sketch
of the self-similar evolution of a two-dimensional Rydberg gas
under the blockade constraint. Blue dots correspond to excited
atoms. As time passes the density of excited atoms increases. This
increase can be exactly "compensated" by increasing the
magnification, {\em i.e.}, reducing the field of view. Thus, under an
appropriate rescaling of the spatial length scale $x$ as a function
of time $t$, such that the product $t\, x^{12+d}$ remains constant,
the system appears static. Note, that this scaling regime is
reached only in the absence of radiate decay (or observation times
much shorter than the inverse Rydberg lifetime) and for densities
at which the average distance between excitations is smaller or
equal than the blockade radius. See refs. \cite{Lesanovsky_2013,Gutierrez15} for further details.}\label{fig:SELFSIMILAR}
\end{figure}

In the case of $\Delta=0$, {\em i.e.}, resonant laser excitation, one encounters the so-called \textit{blockade constraint}. This means  the excitation rate of the second atom is strongly suppressed when its distance to the excited atom is closer than the dissipative blockade radius ($|\mathbf{r}|<R$). Conversely, if $|\mathbf{r}|>R$ the second atom can change its state at the maximum rate $\Omega^2/2\gamma$, {\em i.e.}, it behaves like a quasi free particle. Already this simple constraint gives rise to a highly intricate relaxation dynamics in which the Rydberg gas shows self-similar behaviour. This becomes for instance manifest in the fact that the density of excited atoms $n=\frac{1}{N}\sum_k\langle n_k \rangle$ exhibits a power-law time dependence of the form
\begin{eqnarray}
  n(t)\propto t^\frac{d}{12+d},
\end{eqnarray}
with $d$ being the dimensionality of the system. This scaling behaviour is illustrated in fig. \ref{fig:SELFSIMILAR} in which snapshots of the density of Rydberg atoms in a two-dimensional setting are shown. One sees clearly that by properly adapting the field of view as a function of time the density remains constant, which is a confirmation that the density is not a function that depends separately on space and time but indeed only on a specific combination of both. This is discussed in detail in refs. \cite{Lesanovsky_2013,Gutierrez15}.

We now turn to the situation in which the detuning is positive, $\Delta>0$ (for our case of a positive van der Waals interaction coefficient $C_6$). Here one realises the so-called \emph{facilitation constraint} \cite{Lesanovsky_2014}, which means that the excitation of an atom is strongly enhanced, provided that it is positioned the facilitation radius $r_\mathrm{fac}$. Here the rate of excitation is maximal and given by
\begin{equation}
 \Gamma_\mathrm{fac}=\Omega^2/2\gamma.
\end{equation}
Unfacilitated atoms undergo spontaneous state changes at a rate which is on the order of 
\begin{equation}
\Gamma_\mathrm{spon}\approx(\Omega^2/2\gamma)(r_\mathrm{fac}/R)^{12}=[\Omega/(2\Delta)]^2 2\gamma. 
\end{equation}

\begin{figure}
\includegraphics[width=10 cm]{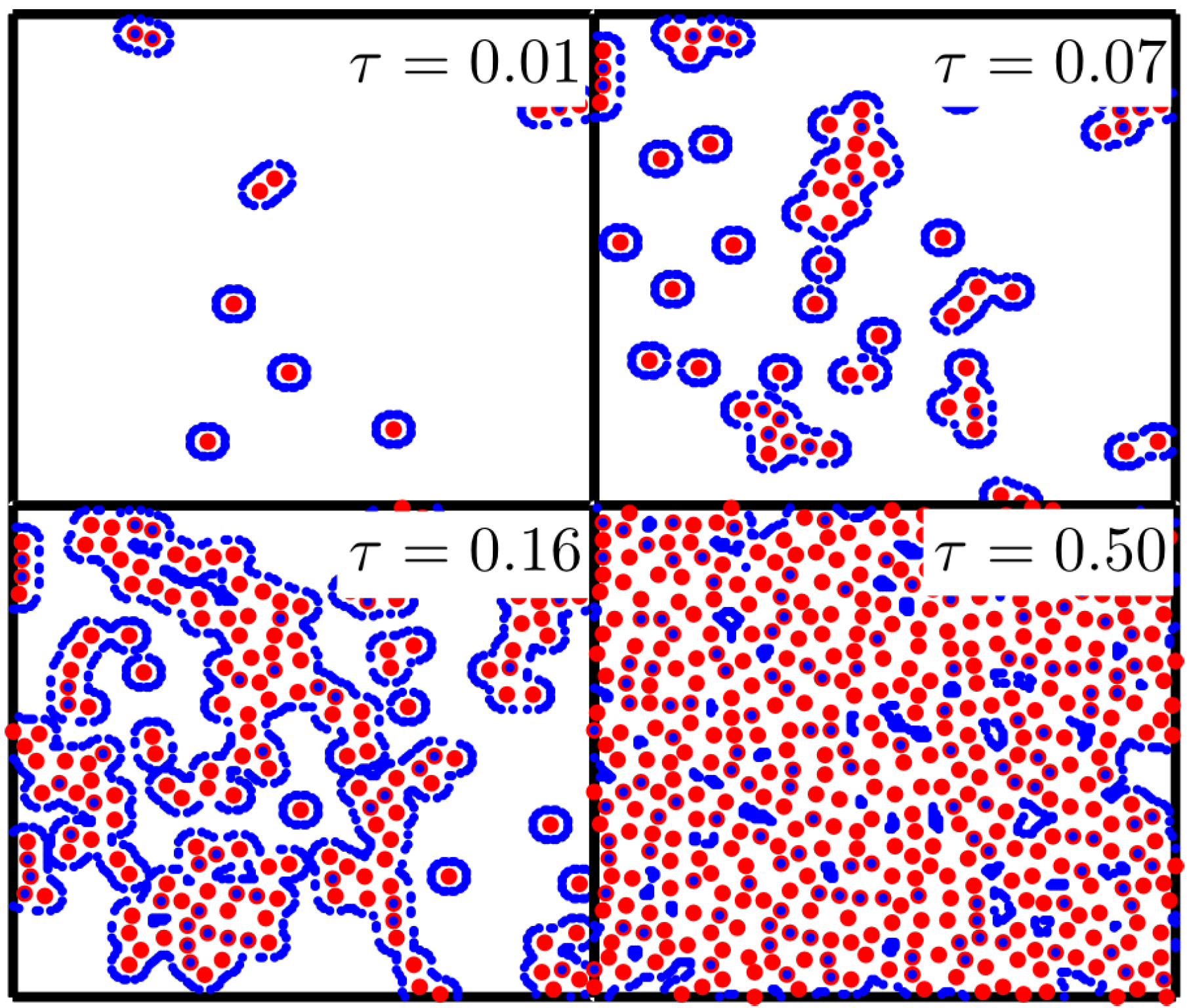}     
\caption{Sketch of
the evolution of a Rydberg gas under the facilitation constraint.
Red dots correspond to excited atoms. Blue atoms are facilitated
atoms whose excitation rate is enhanced. In these simulations the
initial state is devoid of excited atoms and the proliferation of
excitations takes place at the slow rate $\Gamma_\mathrm{spont}$.
Once excitations are created they act as nuclei for the creation of
larger excitation clusters. These clusters grow until they fill up
the available volume. From here onwards facilitated excitation is
strongly suppressed. For details see \cite{Lesanovsky_2014}.}\label{fig:NUCLEATION}
\end{figure}

The dynamics on the many-body level is drastically different compared to the blockade constrained, as is shown in fig. \ref{fig:NUCLEATION}. Starting from a state without initial excitations the first excitation is created at a slow rate $\Gamma_\mathrm{spon}$. This acts subsequently as a nucleus (or "seed") that spawns clusters of excitations to which excitations are added at a rate $\Gamma_\mathrm{fac}$.

\subsection{Kinetic constraints: Experiment}

For an experimental demonstration of the kinetic constraints introduced above, we start by verifying that for our parameters the excitation dynamics is, indeed, in the incoherent regime, so that the approximations of sec. 3.1 can be applied. From an estimate of the intrinsic linewidths of the two excitation lasers and the residual Doppler shift due to the thermal motion of the atoms we find a dephasing rate $\gamma \approx 2\pi\times 700 \,\mathrm{kHz}$, which is also confirmed by the de-excitation experiments reported in sec. 3.4. The dephasing rate of our system is, therefore, expected to be larger than the largest two-photon Rabi frequencies we realize in our experiments, meaning that for all intents and purposes we can neglect the coherent part of the evolution and describe the excitation dynamics by the incoherent spin-flip rate $\Gamma_\mathrm{spon}$ for individual, non-interactiong atoms (eq. 12),
which can take on values up to $280\,\mathrm{kHz}$ for our experimental parameters. Since the experimentally measured lifetime of the $70S$ state is $\tau\approx 80 \,\mathrm{\mu s}$,  the spontaneous decay rate is $\kappa\approx 12.5\,\mathrm{kHz}$ and hence more than an order of magnitude smaller than the excitation and dephasing rates. This separation of timescales allows us to neglect spontaneous decay for now. 

\begin{figure}
\includegraphics[width=14 cm]{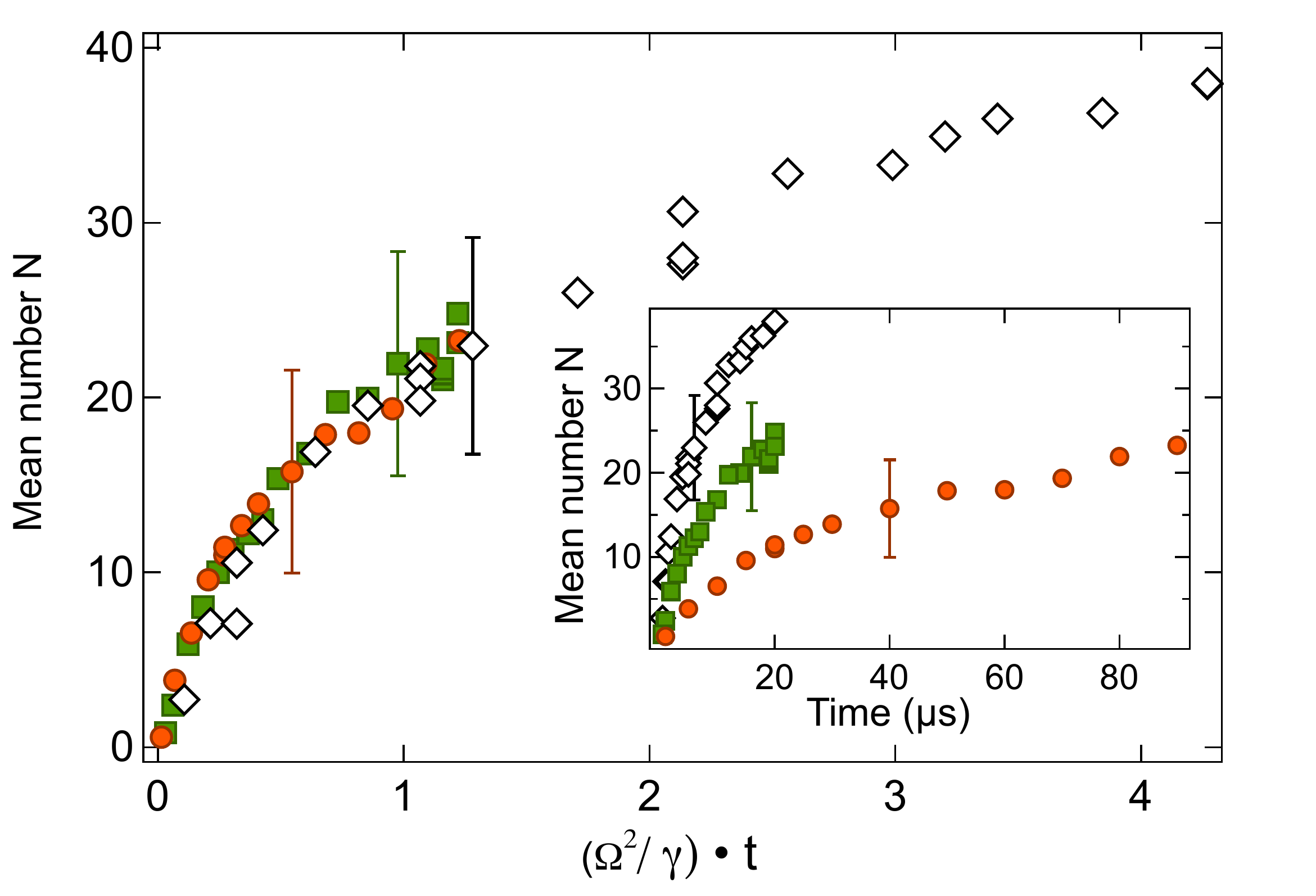}     
\caption{Incoherent excitation of Rydberg states. The inset shows the mean number of excitations as a function of time for three different Rabi frequencies: $\Omega /2\pi = 81$ (open diamonds), $43$ (green squares) and $20$ kHz (red circles). When multiplying the excitation times by $ \Omega^2 / \gamma$, the three curves collapse onto each other (main figure), demonstrating the expected $\Omega^2$ scaling in the incoherent excitation regime. From ref. \cite{Valado_2016}}\label{exp_incoh}
\end{figure}

A simple way of testing whether the excitation dynamics is incoherent is to measure how it scales with the Rabi frequency $\Omega$ \cite{Valado_2016}. For incoherent excitation we expect a scaling with $\Omega^2$~\cite{Urvoy_2015} rather than with $\Omega$ (which is found in the coherent excitation regime~\cite{Heidemann_2007}). To test this, we resonantly excite the $70S$ Rydberg state and measure the average number of excitations as a function of time (see fig. \ref{exp_incoh}). Repeating this experiment for different values of $\Omega$ and then plotting the resulting curves on the same graph, with the horizontal axis scaled in terms of the expected incoherent excitation rate $\propto\Omega^2/\gamma$, we find that the curves collapse on top of each other. From this, we conclude that the excitation dynamics of single Rydberg states is largely incoherent, as expected.

\subsubsection{\em The blockade constraint}

In fig. \ref{exp_incoh}, a tell-tale sign of the expected blockade constraint for resonant excitation is already evident: as time goes on, the slope of the excitation curve diminishes, indicating that the probability per unit time of an atom being excited to a Rydberg state is suppressed as the number of Rydberg excitations grows, and hence an increasing fraction of the interaction volume is excluded from the dynamics through the blockade constraint.

In order to study the blockade constraint more systematically, we excite the atoms resonantly, with $\Delta=0$, and vary the number of atoms per blockade length $\frac{R}{a}$ (where $a=\left(\frac{V_{exc}}{N_g}\right)^{\frac{1}{3}}$ is the mean distance between $N_g$ ground state atoms in the excitation volume $V_{exc}$, and $R= 11.1\,\mathrm{\mu m}$  for our parameters) by changing the effective density of the MOT as described in sec. 2. In this way, we can prepare samples with atom numbers corresponding to values of $\frac{R}{a}$ between around 1.3 ({\em i.e.},  close to the non-interacting case  $\frac{R}{a} < 1$ ) and  $\frac{R}{a}=4.2$.

\begin{figure}
\includegraphics[width=14 cm]{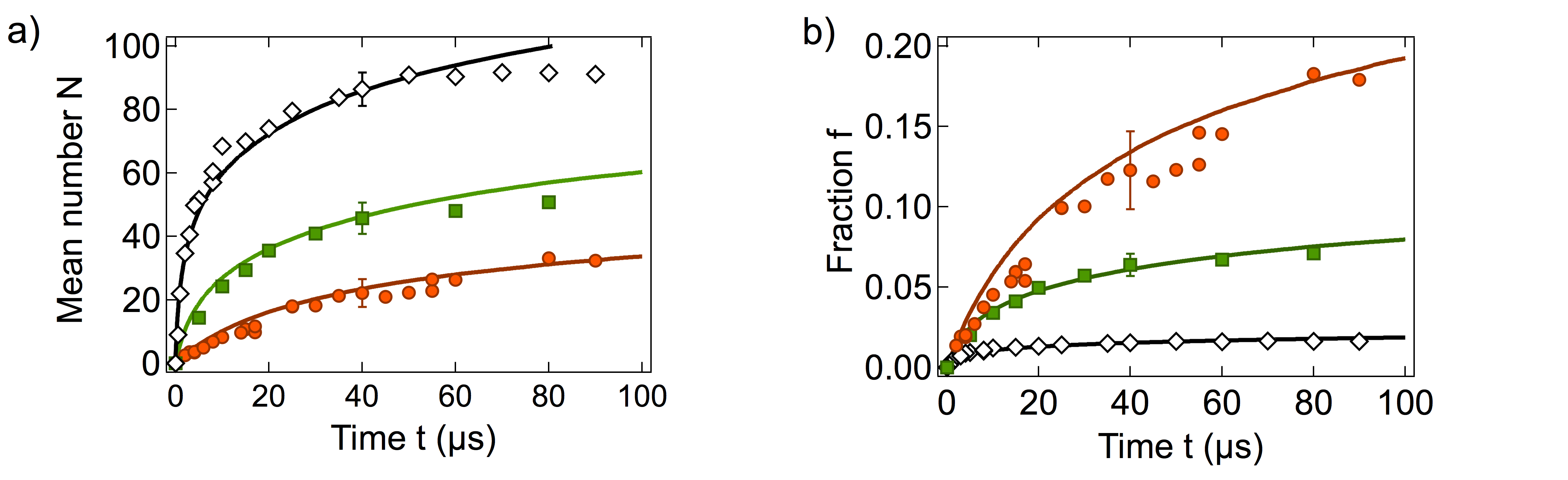}     
\caption{The blockade constraint in a gas of Rydberg atoms. In a) the number of excitations is plotted as a function of time for different ground state atom numbers ($5600$ (open diamonds), $715$ (green squares) and $180$ (red circles)). The blockade constraint is visible as a reduction in the slope of the curves. The solid lines are the results of numerical simulations based on the model described in sec. 3.1. Dividing the data in a) by the respective ground state atom numbers gives the fraction $f$ of excitations plotted in b). In theory, for very long times all the curves are expected to level off at $f=0.5$ (adapted from ref. \cite{Valado_2016}).}\label{exp_blockade1}  
\end{figure}

The results of these experiments  are shown in fig. \ref{exp_blockade1}, together with a numerical simulation based on the theoretical model described above. Qualitative agreement between experiment and theory is excellent, with the cross-over between the initial non-interacting excitation regime (reflected by a linear increase of $N$ with time), and the blockade regime with reduced excitation probability due to interactions clearly visible~\cite{Gutierrez15}. To obtain quantitative agreement we had to scale the theory curves (in $N$) by factors between $0.5$ and $2$. These quantitative discrepancies can be understood from the experimental uncertainties in the measurements of absolute atom numbers, MOT sizes and Rabi frequencies.

The cross-over can also be nicely seen in a plot of the fraction $f=N/N_g$ of excited atoms. In this case, for small numbers of ground state atoms leading to $\frac{R}{a} < 1$ the excited fraction is expected to approach $0.5$, {\em i.e.}, on average half of the atoms are in the excited state. Conversely, for $\frac{R}{a} \gg 1$ a single Rydberg atom inhibits the excitation of a large number of ground state atoms inside a blockade radius, and hence the excited fraction grows much more slowly. In our experiment, $f$ seems to level off around $0.02$ for the largest values of $N_g$, but in theory even in that regime one expects $f$ to reach $0.5$ ({\em i.e.}, the fully mixed state), albeit on extremely long timescales. This is an indication of the glass-like relaxation dynamics due to the blockade constraint, which was referred to in the previous section.

The experimental results of fig. \ref{exp_blockade1} can also be analyzed independently of numerical simulations by calculating the average growth rate of excitations per atom, $(d N/dt)/N_g$, as a function of the average distance $a$ between Rydberg atoms. To this end, the $N$ vs $t$ data from fig. \ref{exp_blockade1} is smoothed (in order to avoid artefacts due to noise), and the growth rate is then extracted by numerical differentiation. Intuitively, one expects that quantity to be essentially constant for $a$ above the blockade radius $R$, for which the excitation events are indepenent, and to decrease sharply below $R$ as the blockade constraint slows down the excitation dynamics.

\begin{figure}
\includegraphics[width=14 cm]{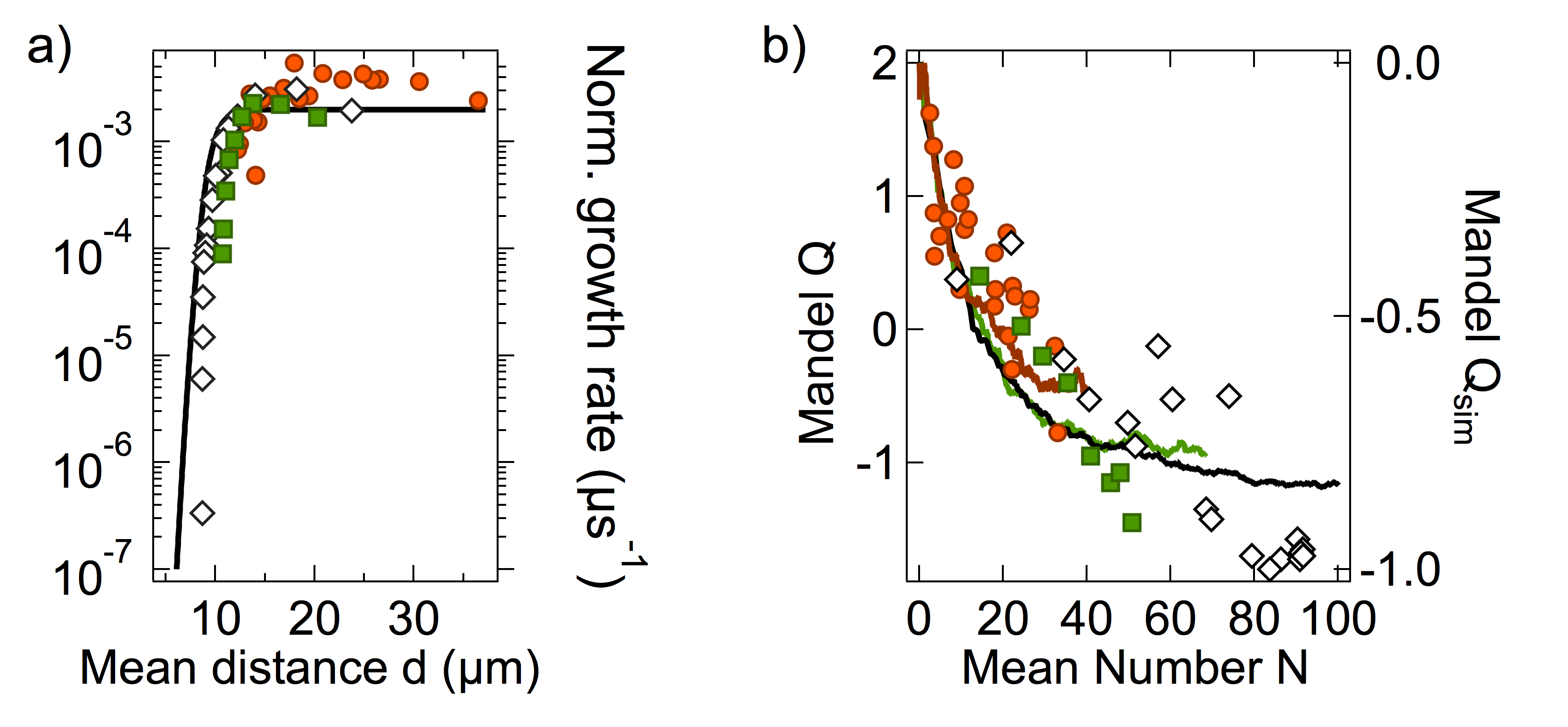}     
\caption{Excitation rate and fluctutations in the blockade constraint. In a) the normalized single-particle excitation or growth rate is plotted as a function of the mean distance $a$ between excited atoms. The dashed line is obtained from the theoretical excitation rate in a mean-field approach. In addition to the reduction of the growth rate, another signature of the blockade constraint is the reduction of fluctuation around the mean number of excitations, reflected in a negative Mandel $Q$-parameter. In b) that parameter is plotted together with the numerical simulation (solid line). Symbols are as in fig. 6. Adapted from ref. \cite{Valado_2016}}\label{exp_blockade2}   
\end{figure}

In fig. \ref{exp_blockade2} , both effects can be clearly seen. In particular, for $d<R$, in the region between $11\,\mathrm{\mu m}$ and $6\,\mathrm{\mu m}$ the growth rate decreases by four orders of magnitude. Contrary to the usual interpretation of the blockade radius (which was originally conceived in the coherent excitation regime \cite{Dudin_2012, Schauss_2012}) indicating a volume in which no more than a single atom can be excited to a Rydberg state, the blockade constraint refers to a drastic slowing down of the dynamics: further excitations can be created, but the excitation rate drops sharply as the distance between the atom to be excited and one or more already excited atoms drops below the blockade radius. In the context of many-body physics this leads to glass-like dynamics and the emergence of the hierarchical, {\em i.e.}, self-similar, relaxation behaviour discussed in sec 3.1. 

Making some reasonable assumptions (as to the number of nearest neighbours in the quasi 1D configuration used in the experiment as well as the distribution of the laser intensity and the atom density inside the MOT) and using eq. 8 (with the mean interparticle distance replacing the sum over individual distances, which corresponds to a mean-field approximation), we can obtain a theoretical prediction for the normalized growth rate per atom, which is plotted in fig. \ref{exp_blockade2} . The main features of the experiment are well reproduced. In particular, with the theoretical curve as a reference point it is evident that the three experimental curves, which were taken for different ground state atom densities, overlap and connect to each other. This highlights the fact that the growth rate only depends on the mean distance between {\em excited} atoms, as it should.

Another way of characterizing the blockade constraint is in terms of correlations. As long as the distance $a$ between excited atoms is much larger than the blockade radius $R$, excitation events are uncorrelated, {\em i.e.}, a new excitation is essentially independent of the instantaenous distribution of excitations inside the system. As $a$ approaches $R$, however, this picture changes. Now, the number of "independent" ground state atoms that do not feel the van der Waals interaction of some already excited atom in the cloud is greatly reduced. Consequently,  the system has fewer choices for distributing additional excitations, which in turn should lead to reduced fluctuations around the mean of the number of excitations. In the non-interacting regime, those fluctuations are Poissonian, whereas in the blockaded regime they are sub-Poissonian, reflecting their correlated character (to emphasize the "exclusion" character of the blockade constraint, we also call this "anti-correlated"). The different characteristics of the fluctuations can be quantified through the Mandel $Q$-parameter~\cite{Viteau_2012} (eq. 5), which by definition is $0$ for perfectly Poissonian statistics and negative for sub-Poissonian fluctuations (which, in the coherent regime, were investigated in~\cite{Gaetan_2009,Urban_2009}).

The experimental results on the $Q$-factor are plotted as a function of the mean number of excitations in fig. \ref{exp_blockade2}. As expected, for large $N$ the $Q$-factor is negative and becomes less negative as $N$ decreases. For $N$ below $\approx 15$ the experimentally measured value of $Q$ is greater than zero, which can be explained by the inevitable additional sources of fluctuations such as laser noise, variations in the atom number, and other experimental imperfections leading to slightly super-Poissonian fluctuations. At the other extreme, for large $N$, the measured $Q$-value drops below $-1$, which is in contrast to the theoretically expected minimum of $-1$ (in the case of vanishing fluctuations). In our experiments, this artefact is due to a possible systematic error in estimating the detection efficiency $\eta$ (by which the experimentally measured values are divided to obtain the actual value of $Q$) as well as possible saturation effects of the detection process and of the peak finding routine, which can yield artificially low values of the variance due to imperfect counting. In spite of these experimental problems, the fact that, again, the three sets of data obtained for different values of $N_g$ collapse onto a single cuve as a function of $N$ shows that $Q$ depends on the number of excited atoms, as expected.

\subsubsection{\em The facilitation constraint}

We now turn to the second type of constraint introduced in sec. 3.1: the facilitation constraint, which occurs for off-resonant excitation with $\Delta>0$ (for our case of repulsive interactions between the $70S$ Rydberg states). In contrast to the blockade constraint, which causes anti-correlations in the dynamics, the facilitation constraint should lead to a strongly (posititvely) correlated evolution \cite{Urvoy_2015, Garttner_2013, Schempp_2014}. In order to explore this regime, we choose a detuning $\Delta/2\pi =+19\, \mathrm{MHz}$, for which $r_{\rm fac} = 6.4\, \mathrm{\mu m}$ and the width of the facilitation shell $\delta r_{\rm fac} = 39\, \mathrm{nm}$. Since we expect the predicted facilitation dynamics to be the more pronounced the larger the overall facilitation volume, which grows with an increasing number of excitations, we choose the 3D configuration for this experiment.

Similarly to the discussion of the blockade constraint, we can qualitatively predict the excitation dynamics by considering the processes expected to occur in the facilitation regime. Most importantly, as at $t=0$ all atoms are in the ground state and, hence, no Rydberg atoms are present, no facilitation events can occur. However, even if off-resonant single particle excitations are suppressed by a factor $\Gamma_{\rm fac}/\Gamma_{spon}=\frac{1}{1+(\Delta/\gamma)^2} \approx1.4 \times 10^{-3}$ compared to the resonant excitation regime, at a certain point a single excitation, also called "seed", will appear in the cloud. At that point, the facilitation mechanism can proceed, creating excitations that, in turn, can facilitate further excitations, and so forth (see fig. \ref{fig:NUCLEATION} ).

Of course, the occurrence of the first excitation is not a deterministic process (contrary to the controlled creation of seed excitations discussed below), and so one does not expect to see a sudden onset of the dynamics at some well-defined time, but rather a slow and gradual start of the dynamics for short times and an acceleration as soon as the probability of there being at least one seed excitation in the cloud approaches unity. The ensuing avalanche-like chain reaction of facilitation events will continue until it reaches the edges of the interaction volume. At that point the excitation dynamics should slow down, and further excitations inside the cloud will be governed by the blockade constraint which, combined with the intrinsic suppression of the excitation rate due to the off-resonant condition, will lead to a dramatic slowing down of the dynamics.

\begin{figure}
\includegraphics[width=14 cm]{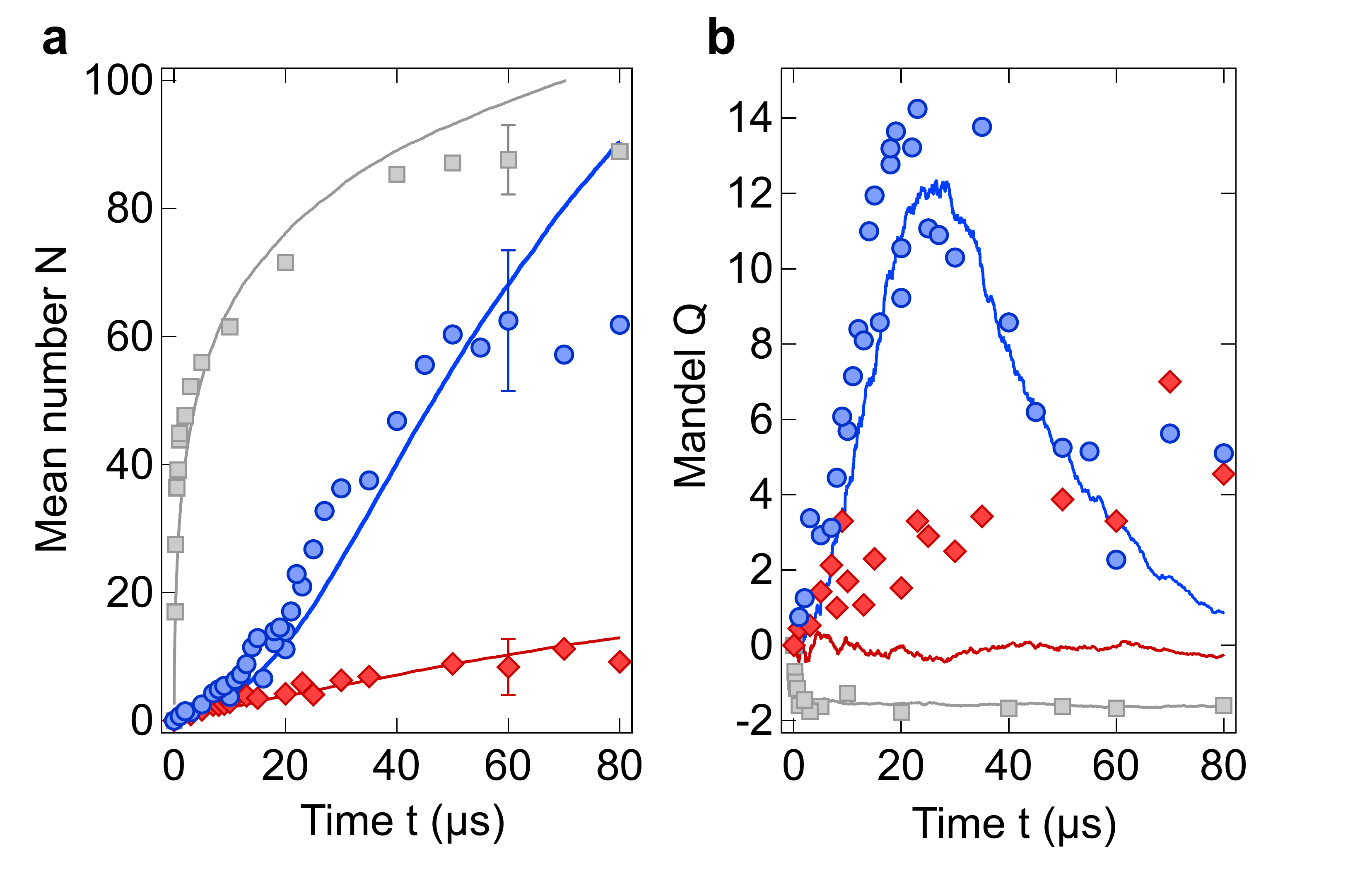}     
\caption{The facilitation constraint in a gas of Rydberg atoms. In a) the mean number of excitations is shown as a function of time for detuning $\Delta/2\pi =+19\, \mathrm{MHz}$ (blue circles), $\Delta/2\pi =-19\, \mathrm{MHz}$ (red circles) and $\Delta=0$ (grey circles).  For positive detuning, the facilitation constraint is evident in the initially slow but then acceleration excitation dynamics (as opposed to the initially fast and then slow dynamics for the blockade constraint with $\Delta=0$). The data for negative detuning underline the importance of the interactions (and their matching to the detuning) for the facilitation constraint. The Mandel $Q$-parameter plotted in b) is further evdience for the facilitation constraint. Adapted from ref. \cite{Valado_2016}.}\label{exp_facilitation1} 
\end{figure}

The experimental results shown in fig. \ref{exp_facilitation1}  confirm that this intuitive picture is correct. There, the three stages of the dynamics can clearly be distinguished: the initial nucleation stage (for $ t< 10\, \mathrm{\mu s}$), in which $N$  grows slowly due to off-resonant single particle excitations; the facilitation stage ($10\, \mathrm{\mu s} <t< 50\, \mathrm{\mu s}$), in which the number of excitations grows increasingly fast due to successive facilitation events starting from an initial seed; and a saturation stage ($t>50\, \mathrm{\mu s}$), in which the dynamics decelerates due to the finite size of the atomic cloud. This regime is visible in the experimental data but already affected  by spontaneous decay, which is not included in the simulations. Apart from that, the experimental results agree well with the numerical simulations.

Similarly to the blockade constraint, also in the case of the facilitation constraint the underlying correlations in the excitation dynamics can be seen in the behaviour of the Mandel $Q$-parameter. Whereas the signature of the blockade constraint is the sub-Poissonian statistics corresponding to a negative $Q$-parameter, the facilitation constraint is reflected in a positive value of $Q$. This can be understood as follows: since the facilitation dynamics is triggered by randomly appearing seed excitations, the fluctuations inherent in the Poissonian statistics of the seed excitations are amplified by the facilitation avalanche triggered by them. The variance expected for such a process is clearly larger than that of a simple single-atom Poissonian excitation event. In our experiments, therefore, we expect to see an increase in $Q$ towards positive values in the facilitation stage, whilst in the saturation stage $Q$ should decrease again as the facilitated dynamics slows down. Experimental results on the Mandel $Q$-parameter are shown in fig. \ref{exp_facilitation1}. As expected from the intuitive picture, $Q$ grows up to $30\, \mathrm{\mu s}$, becoming large and positive, and then decreases again. The fact that for long times $Q$ does not tend to $0$ is, again, a result of additional experimental fluctuations. 

While the above results and intuitive pictures convincingly show the effects of the facilitation constraint, they are slightly complicated by the interplay between spontaneous (off-resonant) seed excitations and facilitated excitations. In the remainder of this section, we demonstrate how the two processes can be studied separately in order to obtain a more complete understanding \cite{Simonelli_2016}.

First, we consider the off-resonant seed excitations. As shown in sec. 3.1, the rate for those excitations is proportional to $\frac{1}{1+(\Delta/\gamma)^2}$ and hence depends on the square of the detuning. Therefore, if we choose the same modulus of the detuning as in the above experiments, but with the opposite sign ({\em i.e.}, negative or red detuning), we expect the excitation dynamics for single off-resonant excitations to be the same, but without the facilitation events, which are absent for negative detuning. Fig. \ref{exp_facilitation1} shows the results for such an experiment, with $\Delta/2\pi =-19 \, \mathrm{MHz}$. Clearly, the dynamics is much slower for long times than in the blue-detuned (facilitation) case, and for $t<10\, \mathrm{\mu s}$ the curves for the two values of $\Delta$ are practically indistinguishable. This is exactly what is expected, as in the nucleation stage only single off-resonant excitations occur, and those do not depend on the sign of the detuning.

\begin{figure}
\includegraphics[width=16 cm]{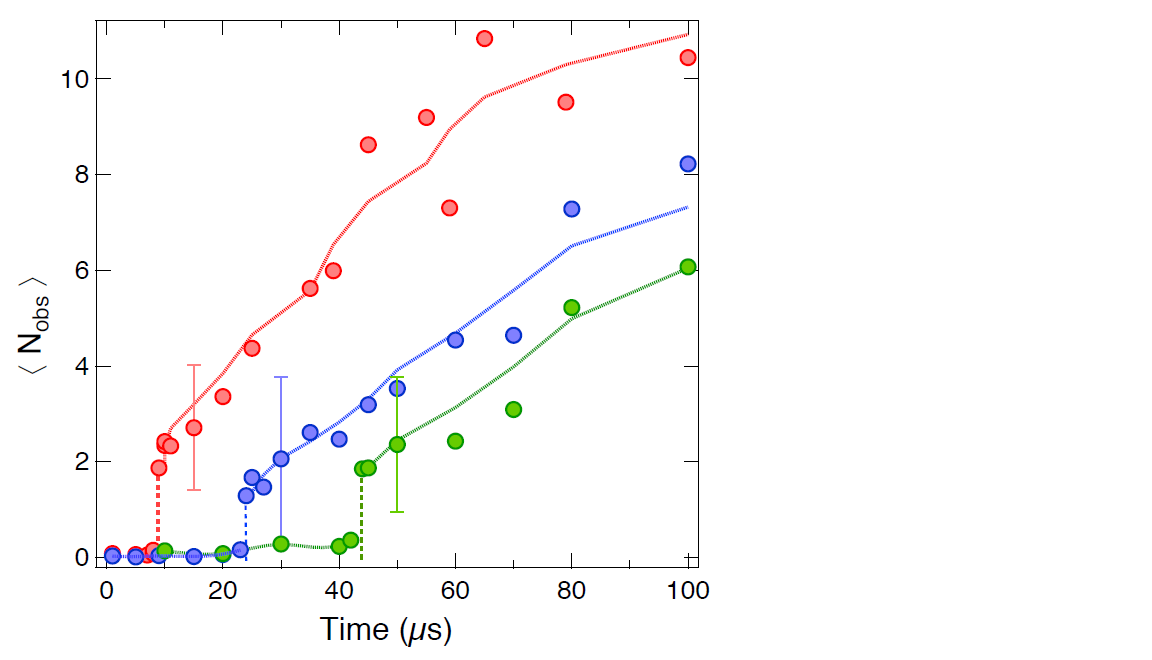}     
\caption{Isolating the facilitation constraint by seeding. This graph shows the mean number $\langle N_{obs}\rangle$ of Rydberg excitations as a function of time for off-resonant excitation at $\Delta/2\pi =+75\,\mathrm{MHz}$ and different values of the time $t_ {seed}$ at which $\approx 2$ seeds are created ($t_ {seed}=10\,\mathrm{\mu s}$ (red), $t_ {seed}=25\,\mathrm{\mu s}$ blue, and $45\,\mathrm{\mu s}$ (green)). The onset of the facilitation dynamics in correspondence with the creation of the seeds is evident. From ref. \cite{Simonelli_2016}.}\label{exp_seed} 
\end{figure}

Second, in order to isolate the first facilitation event, we conduct an experiment similar to the one described above, but with a larger value of $\Delta$ such as to suppress the spontaneous seed excitations as much as possible (here we use $\Delta/2\pi=+75\,\mathrm{MHz}$). Compared to the above experiment, the off-resonant excitation rate is suppressed by a further factor $\approx 15$, meaning that the duration of the nucleation stage is expected to be $\approx 150\, \mathrm{\mu s}$ rather than  $10\, \mathrm{\mu s}$, which is longer than the entire duration of the experiments reported thus far. In order to see any significant dynamics, therefore, we inject seed excitations into the cloud using a short (around $0.5\,\mathrm{\mu s}$) resonant pulse. In fig. \ref{exp_seed}  we show typical experimental results in which around $2$ seed excitations at some finite time $t_ {seed}$. In those experiments, between $t=0$ and  $t=t_ {seed}$ the mean number of excitations grows very slowly at around $10^{-3}\, \mathrm{\mu s}^{-1}$, but for $t>t_ {seed}$ that rate is close to $0.2\, \mathrm{\mu s}^{-1}$, {\em i.e.}, $200$ times higher. This clearly demonstrates that it is the first seed excitation that triggers the avalanche-like facilitation process. It is also evident from fig.  \ref{exp_seed} that the excitation dynamics after the injection of the seed is largely independent of the time at which the seeds are created, as one might expect.

\begin{figure}
\includegraphics[width=12 cm]{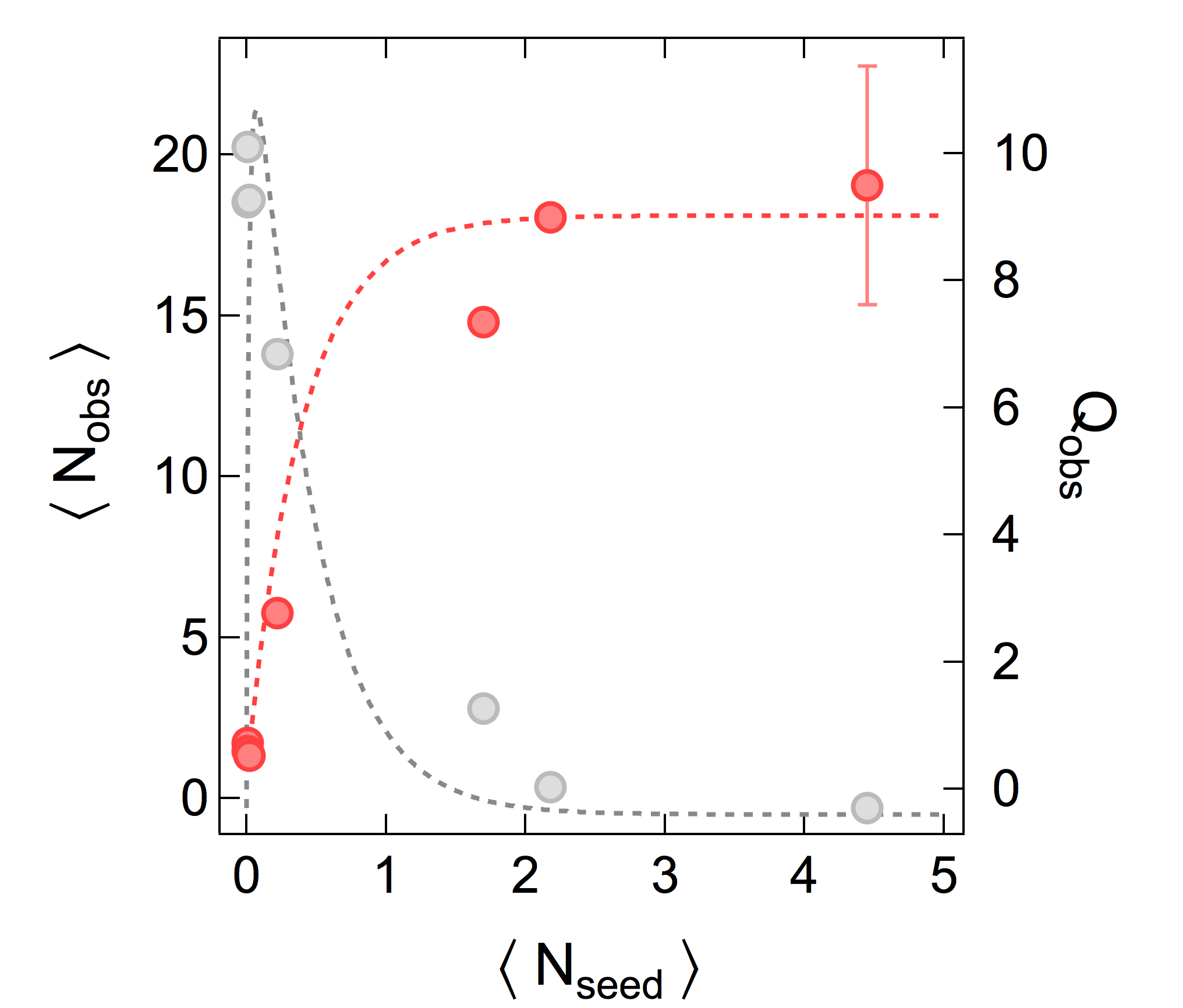}     
\caption{Origin of the super-Poissonian fluctutations in the facilitation constraint. Plotting the the mean number $\langle N_{obs}\rangle$ of Rydberg excitations (red circles and left axis) and the Mandel Q-parameter $Q_{obs}$ (grey circles and right axis) after $100\,\mathrm{\mu s}$ excitation as a function of the number of seed excitations $\langle N_{seed}\rangle$, one observes that while the mean number increases steadily, the $Q$-parameter becomes large and positive between $0$ and $1$ seed excitations. In that regime, the final number of excitations depends on the probability of creating at least one seed excitation at $t=0$. Adapted from ref. \cite{Simonelli_2016}.}\label{exp_seedprob}
\end{figure}

Further confirmation of our interpretation of the role of the seed excitation can be obtained by creating a variable number $\langle N_{seed}\rangle$ of seed excitations at $t=0$ and  then off-resonantly exciting the atoms for $100\,\mathrm{\mu s}$. The mean number of seeds $\langle N_{seed}\rangle$ created at the beginning of the dynamics is varied between 0 and around $5$ by changing the resonant pulse duration. From the results shown in fig. \ref{exp_seed}  one might expect that each time at least one seed is created at $t=0$, a facilitation avalanche is triggered and at $100\,\mathrm{\mu s}$ the system has reached the saturation regime. This behaviour is confirmed by the experimental data of fig. \ref{exp_seedprob}.

We can now also confirm our intuitive picture of the reason for the large and positive $Q$-parameter in the facilitation regime (see above), which we attributed to the amplification of the fluctuations created by the random creation of initial seeds. If we create a small and variable number of seeds at $t=0$, we effectively vary the probability of there being at least one seed excitation at $t=0$. We model our experimental results of fig. \ref{exp_seedprob} using a simple bimodal approach described by the following   probability distribution $P(N)$:
\begin{equation}
P(N)=\alpha\delta(N- N_1) +(1-\alpha)\delta(N-N_2),
\end{equation}
where for a given number of seeds  $\langle N_{seed}\rangle/\eta$,  the quantity  $\alpha$ is the probability of having no seed
\begin{equation}
\alpha=e^{-\frac{\langle N_{seed}\rangle}{\eta}},
\end{equation}

and $N_1$ and $N_2$ represent the number of Rydberg excitations for the two modes of the model. The basic assumption is that in the absence of a seed at $t=0$ the number of excitations in the system will be $N_1\approx 0$, whereas when a seed is created the successive facilitation processes lead to a final number $N_2$ of excitations.
For the above distribution $P(N)$  the  mean number  and the Mandel $Q$-parameter  are
\begin{eqnarray}
\langle N\rangle&= \frac{\langle N_{obs} \rangle}{\eta}= \alpha N_1 +(1-\alpha)N_2, \nonumber \\
Q&=\frac{Q_{obs}}{\eta}=\frac{\alpha (\langle N\rangle- N_1)^2+(1-\alpha)(\langle N \rangle- N_2)^2}{\langle N\rangle}-1.
\end{eqnarray}
\noindent We use these expressions to reproduce the dependence of $\langle N_{obs} \rangle$ and $Q_{obs}$ on $N_{seed}$ reported in fig. \ref{exp_seedprob} by using reasonable values for $N_1$ and $N_2$.  The agreement between the experiment and the model is good. For small values of $\langle N_{seed}\rangle$ the fluctuations in the mean number of excitations at $100\,\mathrm{\mu s}$ are large as the system will sometimes (when a seed is created at $t=0$) end up with a large number of excitations, and sometimes with very few. As the mean number of seeds (and hence the probability of creating at least one seed) grows, the $Q$-parameter decreases and becomes slightly negative, indicating a sub-Poissonian distribution that is compatible with the interpretation of almost deterministically triggering an avalanche that always results in the same final number of excitations.

\begin{figure}
\includegraphics[width=14 cm]{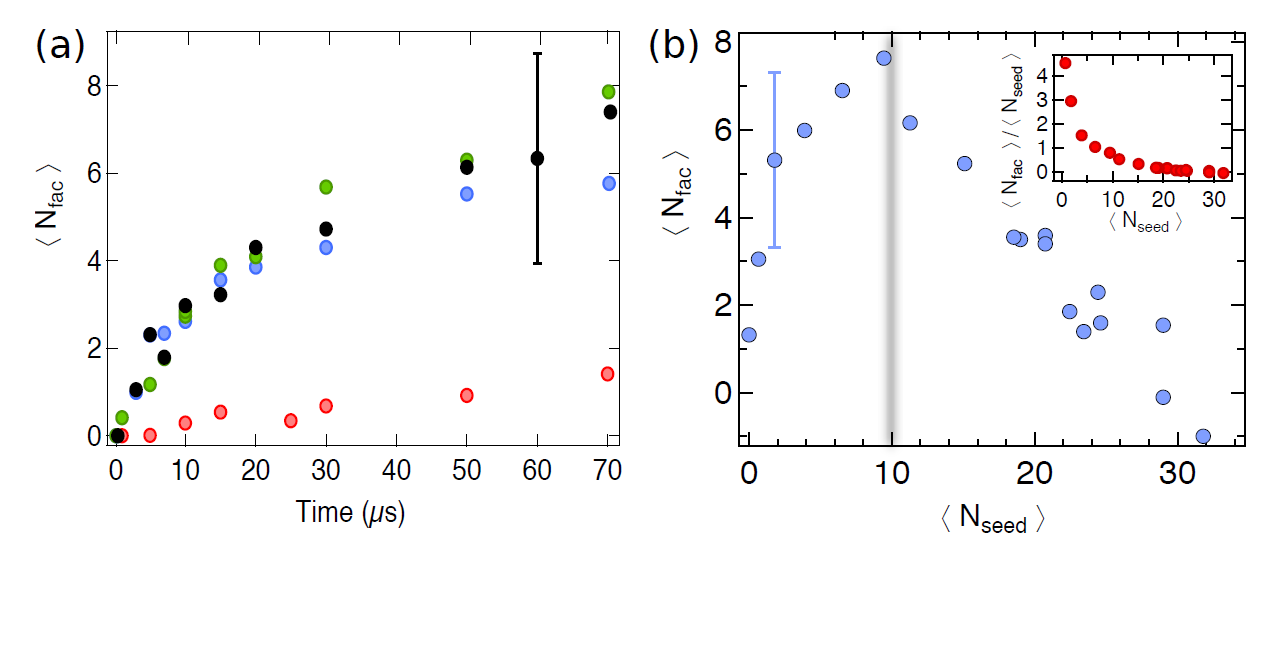}     
\caption{Facilitation dynamics with large numbers of seed excitations. The plot in a) shows the number $\langle N_{\rm fac}\rangle$ of facilitated excitations as a function of the excitation time for different numbers $\langle N_{seed}\rangle$ of initial seeds ($\langle N_{seed}\rangle=1.6$ (green), $\langle N_{seed}\rangle=3.4$ (blue), $\langle N_{seed}\rangle=8.6$ (black), and $\langle N_{seed}\rangle=18$ (red)) with a detuning $\Delta/2\pi=+30 \,\mathrm{MHz}$. In b), $\langle N_{\rm fac}\rangle$ is plotted as a function of $\langle N_{seed}\rangle$ for a fixed excitation time  of $70\,\mathrm{\mu s}$. It is evident that beyond $10$ seed excitations, additional seeds do not increase the number of facilitated excitations. This is also visible in the inset, where the number of facilitated excitations per seed is shown. From ref. \cite{Simonelli_2016}.}\label{exp_seed2} 
\end{figure}

\indent While in fig. \ref{exp_seedprob} the (small) mean number of seeds essentially determined only the probability of starting the facilitation avalanche, in fig.\ref{exp_seed2}  we report results for larger numbers of seed excitations and a fixed excitation time of $70\,\mathrm{\mu s}$. From the above discussion one expects that for large seed numbers each seed will start its own avalanche, up to the point where the seeds are so close together that no further facilitated excitations are possible. This interpretation is confirmed by fig. \ref{exp_seed2} , where $\langle N_{\rm fac}\rangle=\langle N_{obs}\rangle-\langle N_{seed}\rangle$, {\em i.e.}, the number of facilitated excitations, is plotted as a function of the number of seeds. Clearly, $\langle N_{\rm fac}\rangle$ decreases sharply beyond around 10 seed excitations, for which the mean distance between seeds is around $2 r_{\rm fac}$ (for the detuning $\Delta/2\pi=+24 \,\mathrm{MHz}$ used in fig. \ref{exp_seed2} , $ r_{\rm fac}=5.7\,\mathrm{\mu m} $). Plotting the ratio of the number of facilitated excitations and the number of seeds, one finds that for small numbers of seeds (up to about $2$) each seed triggers an avalanche of around $4$ facilitated excitations, whereas above $5$ seed excitations that ratio drops below $1$. Again, this confirms the picture of a large number of seed excitations "getting in the way" of each other and not permitting the onset of a facilitation avalance.

\subsection{Non-equilibrium phase transitions: Theory} 
\begin{figure}
\includegraphics[width=12 cm]{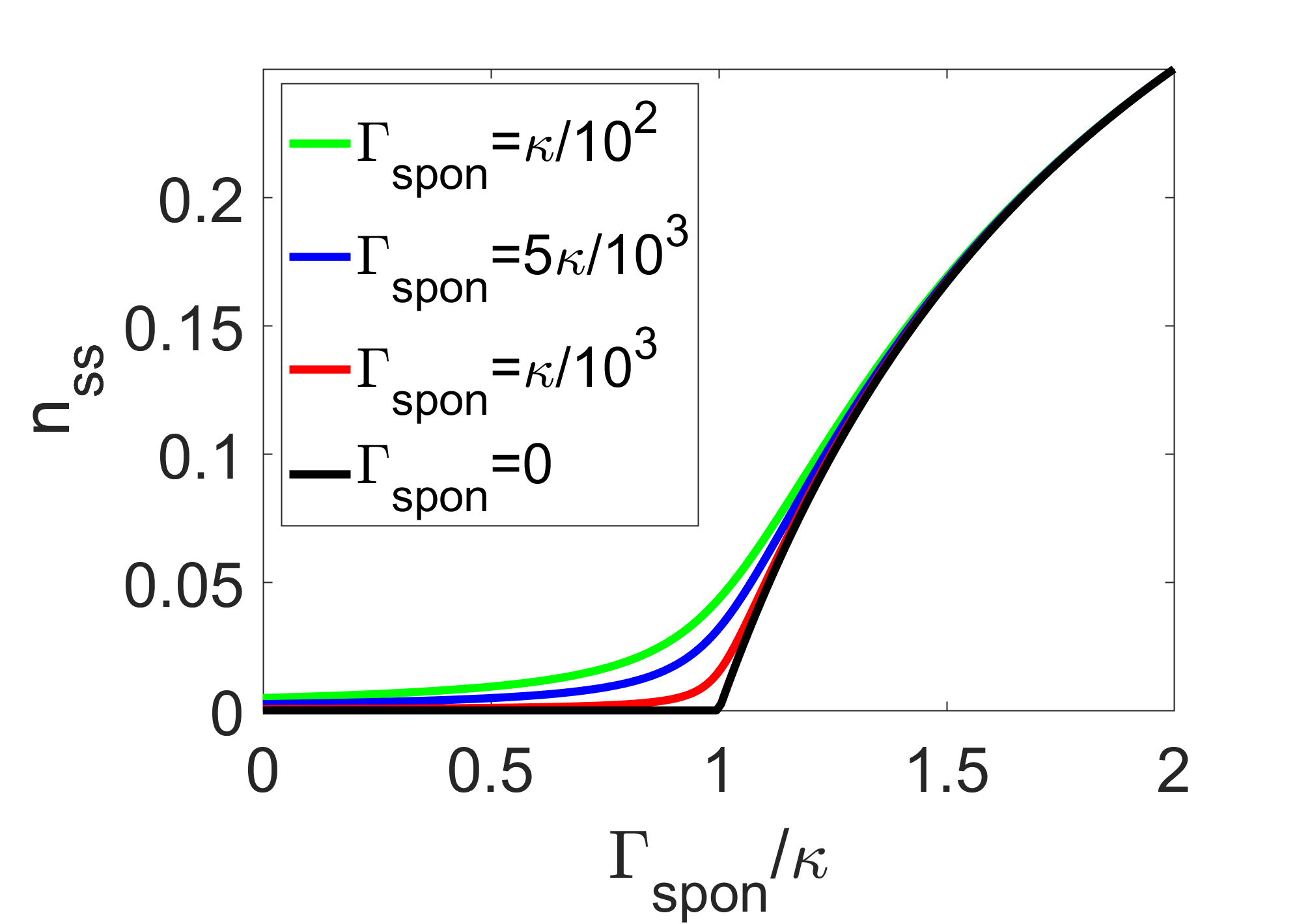}     
\caption{Stationary state solution of the meanfield
equation (\ref{eq:classical_meanfield}). In the absence of spontaneous
(de-)excitation events ($\Gamma_\mathrm{spon}=0$) the system
undergoes a continuous phase transition when the rate of
facilitated excitation, $\Gamma_\mathrm{fac}$, is increased. Below
the transition point the reaches an absorbing stationary state does
not exhibit fluctuations and in which the density is
$n_\mathrm{ss}=0$. The phase transition turns into a cross-over
once $\Gamma_\mathrm{spon}>0$. Note, that numerical simulations
beyond the meanfield approximation suggest that the observed
transition belongs to the directed percolation universality class.
For further discussions see refs. \cite{Hinrichsen_2000,Marcuzzi_2015,Gutierrez_2017}.
}\label{fig:PHASETRANSITION}
\end{figure}

In our discussion of kinetic constraints so far we have neglected radiative decay of Rydberg states back to the ground state level. However, in practice these dissipative processes are always present (in particular at long times) and interestingly their competition with the facilitation constraint leads to an intricate stationary-state behaviour of the Rydberg gas \cite{Marcuzzi_2015,Gutierrez_2017}. To illustrate this we consider a simple one-dimensional lattice model in which we take into account three processes:
\begin{enumerate}
  \item the facilitated (de-)excitation of an atom next to an excited one, at rate $\Gamma_\mathrm{\rm fac}$;
  \item the spontaneous (de-)excitation of atoms, at rate $\Gamma_\mathrm{spon}$;
  \item the radiative de-excitation of an excited atom, at rate $\kappa$.
\end{enumerate}
Combining these processes leads within a meanfield approximation --- in which we assume the system to be homogeneous --- to the following equation for the dynamics of the excitation density $n$:
\begin{eqnarray}
  \frac{\partial n}{\partial t}=\Gamma_\mathrm{fac}n(1-2n)+\Gamma_\mathrm{spon}(1-n)(1-2n)-\kappa n.\label{eq:classical_meanfield}
\end{eqnarray}
We analyse the stationary state of this equation by first considering the limit $\Gamma_\mathrm{spon}/\Gamma_\mathrm{fac}\rightarrow 0$, {\em i.e.}, when only the competition between facilitation and radiate decay governs the dynamics. If $\Gamma_\mathrm{fac}<\kappa$ the stationary state is devoid of excitations, {\em i.e.}, $n_\mathrm{ss}=0$. This is the so-called absorbing phase which is only stable up to the point where $\kappa$ exceeds the value $\Gamma_\mathrm{fac}$. From here onwards the stationary state density becomes finite and reaches the value $n_\mathrm{ss}=\frac{1}{2}\left[1-\frac{\kappa}{\Gamma_\mathrm{fac}}\right]$. The meanfield calculation thus predicts a continuous phase transition between an absorbing phase without excitations and a so-called active phase with a finite density of excitations. This is shown in fig. \ref{fig:PHASETRANSITION}.

Concomitant with a continuous transition is scaling behavior, {\em e.g.}, in the vicinity of the critical value of the control parameter (here $\Gamma_\mathrm{fac}$), one observes a power-law behaviour of the form
\begin{eqnarray}
  n_\mathrm{ss}\sim |\Gamma_\mathrm{fac}-\Gamma^\mathrm{c}_\mathrm{fac}|^\beta\sim
  |\Omega-\Omega_\mathrm{c}|^\beta, \label{eq:scaling_relation}
\end{eqnarray}
where $\beta$ is the so-called static exponent and $\Gamma^\mathrm{crit}_\mathrm{fac}=\gamma$. Obviously, meanfield predicts $\beta=1$ but an exact numerical simulation of the one-dimensional model shows that $\beta\approx 0.27$ \cite{Everest_2016}. This, together with the fact that there is a scalar order parameter (excitation density $n$) and the absence of any apparent symmetries, suggests that the phase transition may in fact belong to the \emph{directed percolation} universality class, see refs. \cite{Hinrichsen_2000}.

The presence of spontaneous excitation processes, parameterised by the rate $\Gamma_\mathrm{spont}$, has a drastic impact on the physics. It removes the absorbing state, {\em i.e.}, the solution $n=0$. As shown in fig. \ref{fig:PHASETRANSITION} this leads to a smoothing of the phase transition which is rendered into a crossover \cite{Gutierrez_2017}. Nevertheless, for sufficiently weak $\Gamma_\mathrm{spont}$ one still expects the occurrence of a scaling region in which one can observe behaviour of the form (\ref{eq:scaling_relation}).

\subsection{Non-equilibrium phase transitions: Experiment}

In sec. 3.2 we experimentally demonstrated the facilitation process and the controlled creation of seed excitations in our system. In the language of absorbing state phase transitions introduced above, the facilitation process corresponds to offspring production, which is one of the ingredients needed to physically implement a model for an absorbing state phase transition. The creation of seed excitations, on the other hand, is a prerequisite for studying the phase transition itself: without such a process, the system would remain in the absorbing state (= all atoms in the ground state) forever, and the regions in the phase diagram with a finite fraction of excited atoms could not be studied.

Finally, in order to fully implement a model system exhibiting an absorbing state phase transition, we need a dissipative process corresponding to the "sudden death" of an excited atom. One way of implementing that process is through spontaneous decay of the Rydberg states, as introduced in the previous section. Later in this section we will also study a mechanism for induced dissipation through de-excitation of excited atoms.

To obtain some experimental insight into the behaviour of our experimental "driven-dissipative" model system for an absorbing state phase transition, we study the stationary state of our system as a function of the two control parameters $\Delta$, which controls $r_{\rm fac}$ and $\delta r_{\rm fac}$,  and $\Omega$ \cite{Gutierrez_2017}. The protocol for this is as follows. At the beginning of an experimental cycle (during which the MOT beams are switched off), we excite around $6$ seed excitations (according to a Poissonian distribution) in $0.3\, \mu\mathrm{s}$ with the excitation laser on resonance with the Rydberg transition. Thereafter, the atoms are excited at finite detuning $\Delta > 0$ and Rabi frequency $\Omega$ for a duration of 1.5 ms, which is around $10$ times longer than the lifetime of the 70S state. The procedure is repeated 100 times for each set of parameters, with a repetition rate of 4 Hz, in order to get reliable estimates of the mean $N_I$ and the variance $\Delta N_I {}^2$ of the number of detected ions.

\begin{figure}
\includegraphics[width=14 cm]{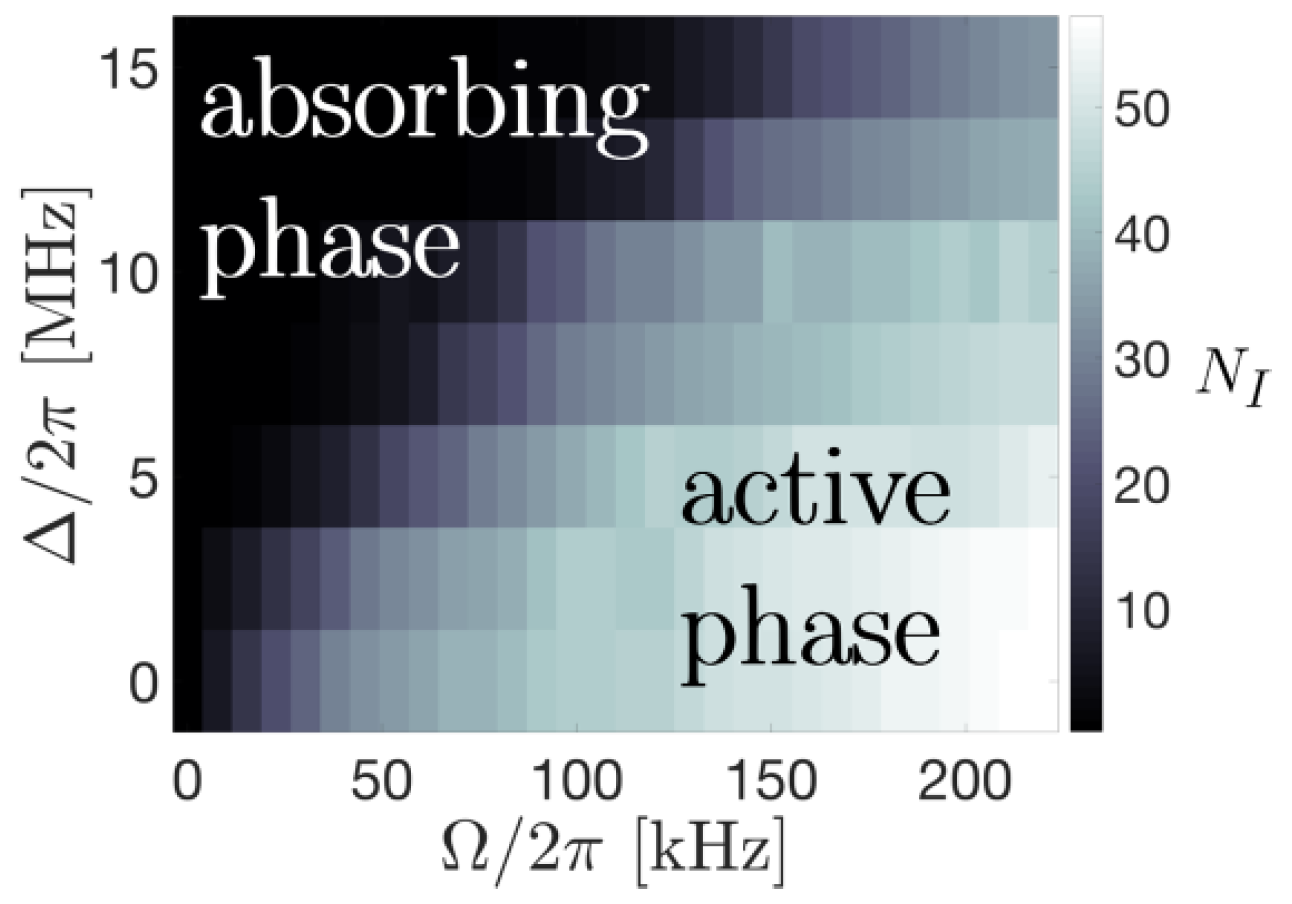}     
\caption{Experimental phase diagram for a driven-dissipative Rydberg gas exhibing an absorbing state phase transition. Plotted here as a function of $\Omega$ and $\Delta$ is the number of excitations after $1.5\, \mu\mathrm{ms}$ off-resonant excitation of a gas prepared with $6$ seed excitations. The crossover between the absorbing phase with $0$ excitations and the active phase with up to $50$ excitations is clearly visible. Adapted from ref. \cite{Gutierrez_2017}. }\label{exp_absorbing_phase}   
\end{figure}

The resulting phase diagram is plotted in fig. \ref{exp_absorbing_phase}. The crossover between the absorbing phase with essentially zero excitations for sufficiently small $\Omega$, and an active phase with a finite number of excitations for larger $\Omega$ can be clearly seen. The point of this crossover depends on $\Delta$, with larger $\Delta$ corresponding to a larger value of $\Omega$ for the crossover. This dependence is due to the interplay between different effects. First, the creation of spontaneous seed excitations scales as $1/\Delta^2$, and hence we expect to see the offset of the critical point discussed in the previous section. Second, the width of the facilitation shell and hence the probability of finding an excitable ground state atom inside it depends on $\Delta$, with larger $\Delta$ resulting in a smaller probability. This, together with the effect of the thermal motion of the atoms, results in the critical value of $\Omega$ increasing with $\Delta$ and, eventually, diverging. In practice, this means that in order to realize as "clean" a realization of the absorbing state phase transition as possible, we need to choose a value of $\Delta$ that results in a compromise between those two trends - larger $\Delta$ meaning fewer spontaneous seed excitations that "hide" the critical point, but also reducing the probability of establishing the long-range correlations associated with the active phase.

\begin{figure}
\includegraphics[width=14 cm]{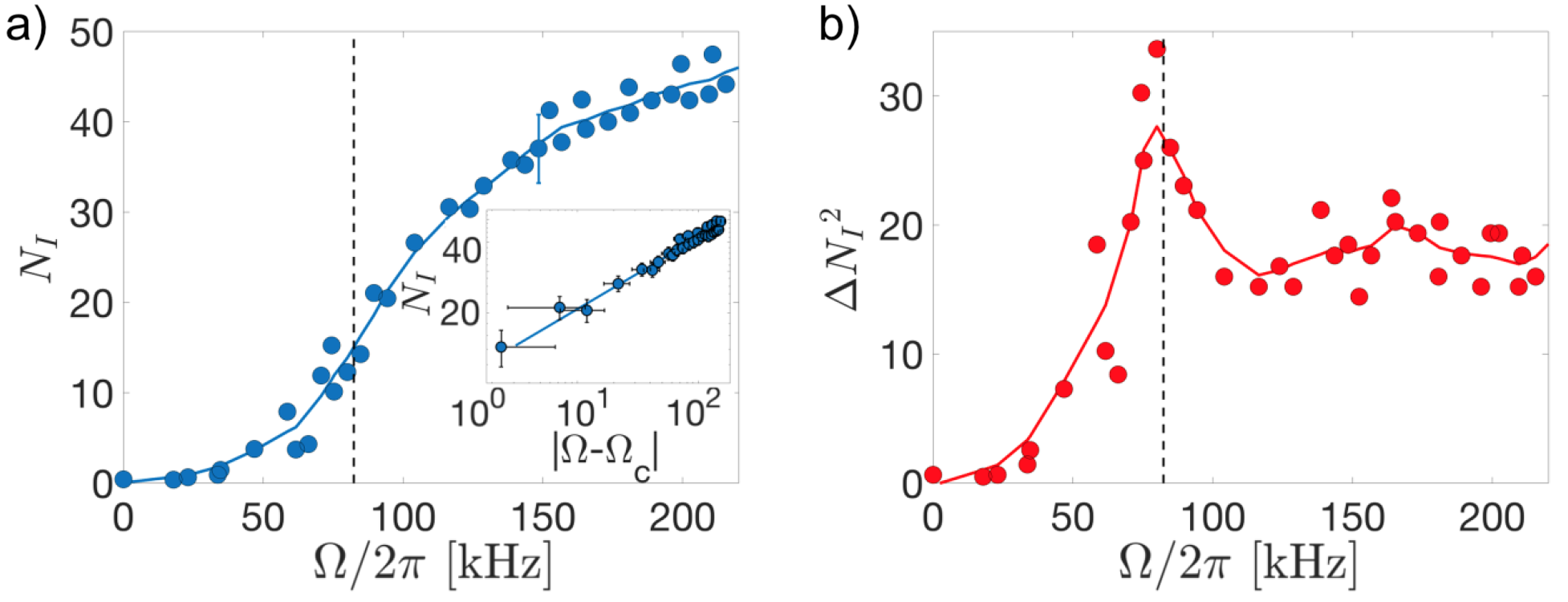}     
\caption{Evidence for an absorbing state phase transition in a Rydberg gas. In a) the number of excitations in the stationary state is plotted as a function of $\Omega$ (the solid line is a sliding average to guide the eye). The inset shows a power-law fit around the critical value $\Omega_c$. In b) the peak in the variance plotted as a function of $\Omega$ indicates the critical point; its position coincides with the value found from the fit in a) (dashed vertical line). Adapted from ref. \cite{Gutierrez_2017}.}\label{exp_absorbing_results}
\end{figure}

For our experimental conditions, we find a detuning $\Delta=2\pi \times 10\,\mathrm{MHz}$ to be a good compromise and choose that value for a more in-depth study of the phase transition. Again, we seed the system with around $6$ excitations and measure the steady-state number of excitations as a function of $\Omega$, obtaining the curve shown in fig. \ref{exp_absorbing_results}. Up to around $\Omega=2\pi\times 30\,\mathrm{kHz}$ the number of excitations is close to zero, indicating that the system is in the absorbing phase. Between $\Omega=2\pi\times 30\,\mathrm{kHz}$ and $\Omega=2\pi\times 100\,\mathrm{kHz}$ the number of excitations increases rapidly, after which the rate of increase diminishes as the system reaches the fully active phase.

Qualitatively, the experimental curve bears a strong resemblance to the theoretical prediction shown in fig. \ref{exp_absorbing_results}. We can also make this comparison more quantitative by extracting the critical exponent $\beta$ from the data (see sec. 3.3). To do so, we pick a probable value for the critical driving strength $\Omega_{c}$ and fit a power-law curve to $N$ as a function of  $|\Omega-\Omega_{c}|$ (see eq. 18). We then optimize $\Omega_{c}$ by maximizing the goodness of that fit, thus obtaining the curve shown in fig. \ref{exp_absorbing_results}. This procedure yields  $\Omega_{c}\approx 2\pi\times 80\,\mathrm{kHz}$, which corresponds roughly to the inflection point of the experimental curve, and the exponent $\beta\approx 0.31$ of the power law fit agrees well with the critical exponent expected for 1D direction percolation.

Another clear sign of the phase transition occurring is an increase in the fluctuations around the mean of the number of excitations. At the critical point the correlations in the system, and consequently also the fluctuations, diverge for an infinite system size. In practice, for a finite system such as ours we expect a maximum of the fluctuations at the critical point. Fig. \ref{exp_absorbing_results} shows this clearly: at $\Omega\approx 2\pi\times 80\,\mathrm{kHz}$, there is a peak in the fluctuations plotted against $\Omega$. Above that critical value for $\Omega$, which coincides with the $\Omega_{c}$ found above by optimizing the power-law fit, the fluctuations decrease slightly and level off at a fixed value beyond  $\Omega\approx 2\pi\times 120\,\mathrm{kHz}$

\begin{figure}
\includegraphics[width=10 cm]{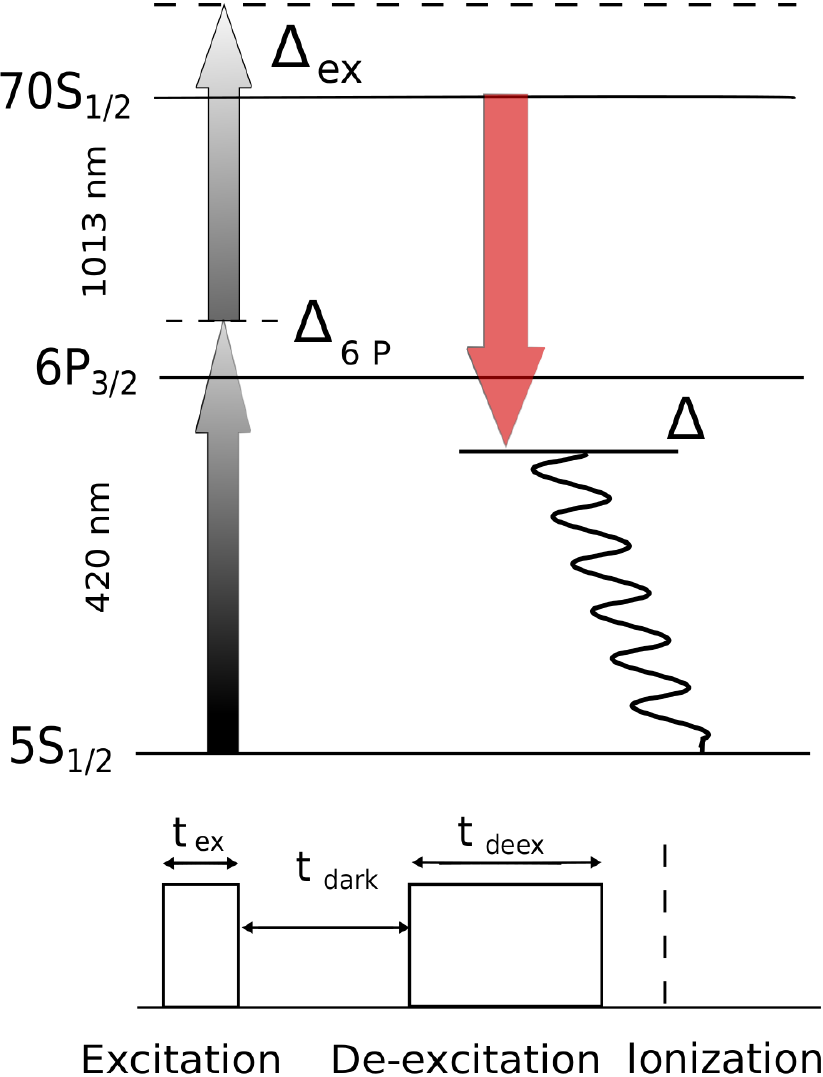}     
\caption{Protocol for controlled deexcitation. After the usual Rydberg excitation with detuning $\Delta_{\rm ex} $, the laser at 1013 nm is tuned into resonance (to within a variable detuning $\Delta$) with the fast-decaying $6P$ state. From ref. \cite{Simonelli_2017}}\label{exp_deex}  
\end{figure}

\subsection{Controlled dissipation through Rydberg de-pumping}

In the above experiments we relied on spontaneous decay to provide the necessary de-excitation mechanism for the absorbing state phase transition. While this approach yielded good results, it still has some shortcomings. Firstly, decay back to the ground state (which can also occur to the non-coupled $F=1$ level of the ground state) is not the only decay channel. Rather, the population of nearby Rydberg states by absorption or emission of black-body photons can significantly complicate the picture \cite{Cooke_1980}.  Secondly, the timescale for spontaneous decay is fixed. It would be useful, however, to be able to reduce that timescale in order to shorten the overall time needed to reach the stationary state, thus avoiding issues related to the thermal motion of the atoms and other mechanical effects due to, {\em e.g.}, the van der Waals repulsion ref. \cite{Faoro_2016}. 

One possible method to artificially shorten the lifetime of the Rydberg state is to actively de-excite (or de-pump) the Rydberg state via a fast decaying intermediate state. We tested such a method (see fig. \ref{exp_deex}  ) \cite{Simonelli_2017}, in wich initially an excitation pulse of duration $t_{ex}$ is applied with both lasers, where the two-photon excitation is detuned by $\Delta_{ex}$ from resonance (the MOT beams are switched off during the entire excitation and de-excitation sequence). After a variable dark time $t_{dark}$, during which both lasers are switched off, only the $1013 \,\mathrm{nm}$  laser is switched on for $t_{deex}$, with the AOM frequency set to a value that shifts the frequency of that laser to be resonant with the transition $70S_{1/2}-6P_{3/2}$ to within a detuning $\Delta$. The $6P_{3/2}$ state has a lifetime $\tau_{6P} \approx 120  \,\mathrm{ns}$. Finally, $300 \,\mathrm{ns}$ after the de-excitation pulse the number of Rydberg excitations is measured.

\begin{figure}
\includegraphics[width=16 cm]{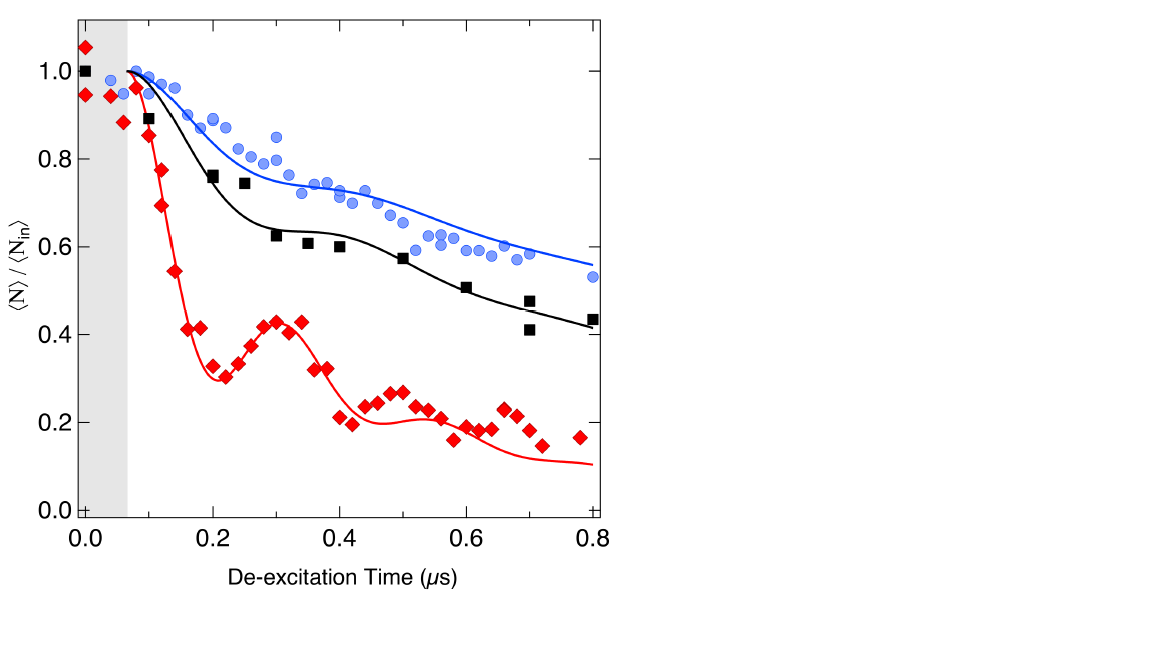}     
\caption{Deexcitation dynamics from the $70S$ Rydberg state. Plotted here is the normalized population of the Rydberg state as a function of the de-excitation time $t_{deex}$ for different de-excitation Rabi frequencies $\Omega_{1013}$ ($\Omega_{1013} \approx 2 \pi \times4\,\mathrm{MHz}$ (red diamonds), $\Omega_{1013}$  ($2 \pi \times 2\,\mathrm{MHz}$, black squares, and $2 \pi \times1.4\,\mathrm{MHz}$, blue circles). The experimental data are well reproduced by a simple damped two-level model (solid lines). From ref. \cite{Simonelli_2017}.}\label{exp_deex_dyn} 
\end{figure}

We first illustrate the de-excitation technique in the non-interacting regime, {\em i.e.}, for sufficiently low Rydberg densities such that the van der Waals interaction can be neglected. As we shall see later, that interaction can significantly affect the de-excitation dynamics . Fig. \ref{exp_deex_dyn}  shows the fraction $\langle N\rangle /\langle N_{in}\rangle$ of Rydberg atoms remaining after the excitation pulse as a function of $t_{deex}$ for different values of $\Omega_{1013}$. For large values of $\Omega_{1013}$ around $2\pi \times 4  \,\mathrm{MHz}$ (which is larger than the frequency associated with decay from the $6 P_{3/2}$ state, 1/$\tau_{6P} \approx 2 \pi \times 1.3 \,\mathrm{MHz}$), the dynamics shows signs of residual coherent oscillations (which we use to calibrate the Rabi frequency of the 1013 nm transition), whereas  for $\Omega_{1013}$ below $ 2 \pi \times 2 \,\mathrm{MHz}$ those oscillations are strongly damped and the de-excitation dynamics can, to a good approximation, be described by an exponential decay. Fig. \ref{exp_deex_dyn} also shows the results of a numerical integration of a simple (coherent) two-level system with a loss term from the $6 P_{3/2}$  state \cite{Shore_2006} . If the de-excitation Rabi frequency is sufficiently small ($\Omega_{\rm 1013}<\gamma$), then similarly to the excitation process discuss in sec. 2, the single-atom de-excitation dynamics can always be described by a rate equation with $\Gamma= \frac{\Omega_{1013}^2}{2 \gamma}\cdot\frac{1}{|1+(\Delta/\gamma)^2|}$.

\begin{figure}
\includegraphics[width=14 cm]{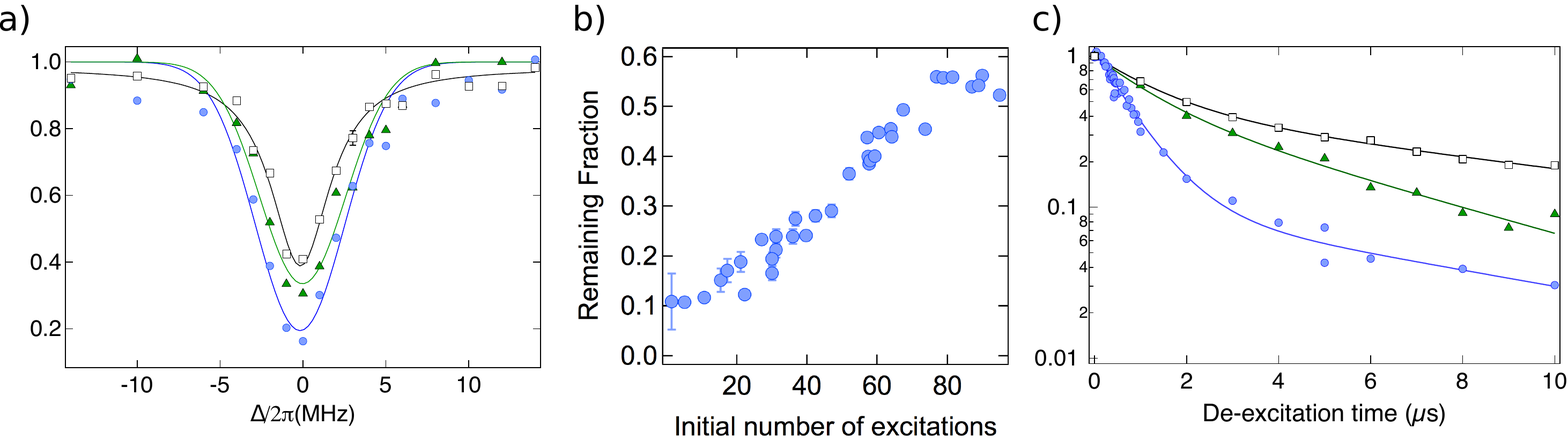}     
\caption{ De-excitation process following resonant excitation for different initial mean numbers $\langle N_{in}\rangle$:  25 (blues circles), 34 (green triangles) and 50 (white squares). The different values of $\langle N_{in}\rangle$ (ranging from the non-interacting to the interacting regime) are obtained by varying $t_{ex}$ between $0.5  \,\mathrm{\mu s}$ and $5  \,\mathrm{\mu s}$. In (a), the remaining fraction of Rydberg atoms $\langle N \rangle /\langle N_{in}\rangle$ is plotted as a function of the de-excitation detuning $\Delta$. Here, $t_{dark} = 0.5  \,\mathrm{\mu s}$ and $t_{deex} = 2   \,\mathrm{\mu s}$. The solid lines are Lorentzian fits to guide the eye. The expected shift in the de-excitation detuning is visible mainly as an increase in the remaining fraction at $\Delta=0$, shown systematically in (b). Here, the remaining fraction after a de-excitation pulse of duration $t_{deex} = 1 \,\mathrm{\mu s}$ is plotted as a function of $\langle N_{in}\rangle$ . In (c), the de-excitation dynamics is shown for $\Delta=0$. Adapted from ref. \cite{Simonelli_2017}.}\label{exp_deex_res}  
\end{figure}

We now proceed to systematically study the de-excitation dynamics. As shown above, in the non-interacting (and incoherent) regime the dynamics can be described by a rate equation, leading to an exponential decrease of $\langle N\rangle $ with $t_{deex} $. For resonant excitation one, therefore, expects to see a minimum in the remaining fraction of Rydberg excitations after the de-excitation pulse as a function of the detuning $\Delta$ for $\Delta=0$. This is confirmed in fig. \ref{exp_deex_res} (a), where for an initial $\langle N_{in}\rangle = 20$, $\langle N\rangle /\langle N_{in}\rangle$ is plotted as a function of $\Delta$ for a fixed $t_{deex} = 2  \,\mathrm{\mu s}$ and $\Omega_{1013} = 2\pi \times 1 \,\mathrm{MHz}$. When $\langle N_{in}\rangle$ is increased to $50$, for which the van der Waals interaction is expected to be non-negligible, the remaining fraction at $\Delta=0$ also increases, indicating that the interactions shift the Rydberg levels and hence the de-excitation laser is no longer resonant.

This effect is shown more systematically in Fig. \ref{exp_deex_res} (b), where the remaining fraction of Rydberg excitations at $\Delta=0$ is plotted as a function of $\langle N_{in}\rangle$. Between the non-interacting regime ($\langle N_{in}\rangle \approx 2$, corresponding to an interatomic distance $a \approx 70 \,\mathrm{\mu m} > R$) and the strongly interacting regime ($\langle N_{in}\rangle \approx 80$, for which $a \approx 2 \,\mathrm{\mu m} < R$), the remaining fraction increases from $0.1$ to $0.6$. This crossover from the non-interacting to the interacting regime is also visible in the de-excitation dynamics. Fig. \ref{exp_deex_res} (c) shows the remaining fraction as a function of $t_{deex}$ for different values of $\langle N_{in}\rangle$. The de-excitation rate decreases appreciably (by up to a factor 6) as $\langle N_{in}\rangle$ is increased. The effect of the van der Waals interactions is also reflected in the fact that the dynamics of the remaining fraction does not follow a simple exponential decay. Rather, the rate of the exponential decay decreases as $t_{deex}$ is increased. We interpret this as a consequence of the spread of inter-atomic distances between the excited atoms, which means that Rydberg atoms with more distant neighbours are de-excited faster, whereas those interacting more strongly with their closer neighbours exhibit reduced de-excitation rates. 

\begin{figure}
\includegraphics[width=14 cm]{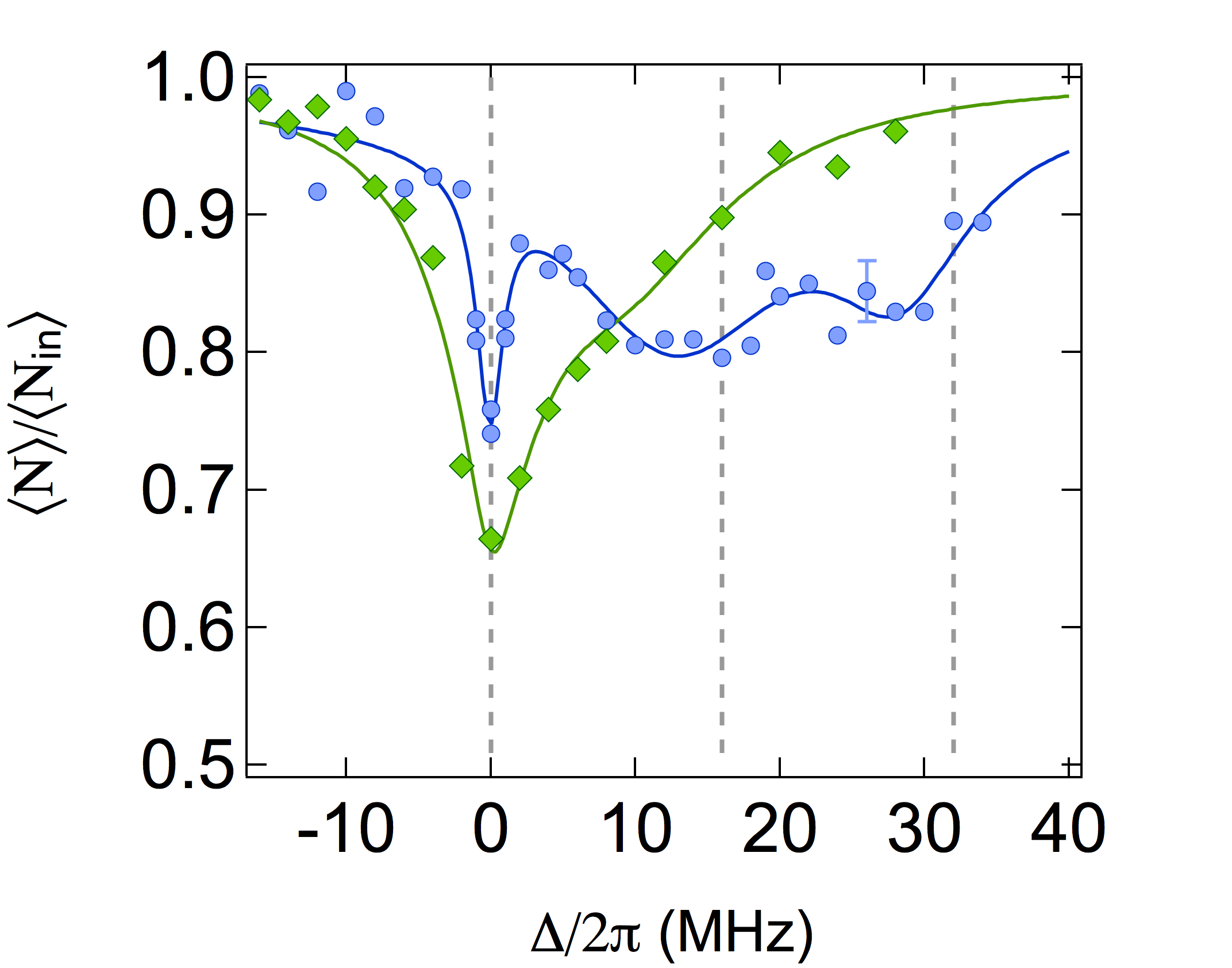}     
\caption{De-excitation process following off-resonant excitation in the facilitation regime. The remaining fraction  $\langle N \rangle /\langle N_{in}\rangle$  after excitation of $\langle N_{in}\rangle \approx 20$ excitations ($t_{ex} = 5 \,\mathrm{\mu s}$) at $\Delta_{ex} = 2\pi \times 16  \,\mathrm{MHz}$ is plotted as a function of $\Delta$. The blue circles correspond to de-excitation ($t_{deex} = 2  \,\mathrm{\mu s}$) after a dark time $t_{dark} = 0.5  \,\mathrm{\mu s}$, whereas the green diamonds are obtained for $t_{dark} = 5  \,\mathrm{\mu s}$. The solid lines are triple-Lorentzian fits to guide the eye. Adapted from ref. \cite{Simonelli_2017}.}\label{exp_deex_fac} 
\end{figure}

The effect of the interactions on the de-excitation process is even more evident if we initially create excitations by the facilitation mechanism, {\em i.e.}, with positive detuning. In that case, atoms excited by that mechanism have an interaction energy with their respective seed atoms that, by definition, is equal to the detuning. In the quasi one-dimensional geometry used in these experiment, this process results in a chain of excitations at a fixed spacing $r_{\rm fac}$. Atoms at the edges of this chain have a single neighbour and hence an interaction energy $\hbar \Delta_{ex}$, whereas atoms inside the chain have two neighbours and a resulting energy shift of $2\hbar \Delta_{ex}$. The de-excitation resonances for those two classes of atoms should, therefore, be centred around $\Delta=\Delta_{ex}$ and $\Delta=2\Delta_{ex}$, respectively. Furthermore, due to the residual thermal motion of the atoms (and also the van der Waals repulsion \cite{Thaicharoen_2015, Teixeira_2015, Faoro_2016}) the distances between the atoms will increase over time, so that eventually each atom will have a de-excitation resonance at $\Delta=0$ as the interactions decrease.

We test the above picture by off-resonantly exciting around $\langle N_{in}\rangle = 20$ atoms at $\Delta_{ex}=2\pi \times 16   \,\mathrm{MHz}$ using a $5  \,\mathrm{\mu s}$ excitation pulse. As the $70S_{1/2}$ Rydberg state used here interact repulsively, the facilitation condition correspond to a positive detuning.  A de-excitation pulse of duration $t_{deex}=2  \,\mathrm{\mu s}$ follows after two different values of a dark time: $t_{dark} = 0.5  \,\mathrm{\mu s}$ and $t_{dark} = 5  \,\mathrm{\mu s}$. In Fig. \ref{exp_deex_fac}, for $t_{dark} = 0.5  \,\mathrm{\mu s}$ three de-excitation resonances can be seen, corresponding to atoms with two neighbours at distance $r_{\rm fac}$ ($\Delta=2\Delta_{ex}$), with one neighbour ($\Delta=\Delta_{ex}$), and without any neighbours ($\Delta=0$ ). The latter class of atoms corresponds to single off-resonantly excited Rydberg atoms that did not lead to further facilitation events, or else to atoms whose neighbours at $r_{\rm fac}$ have already moved sufficiently so as to reduce the interaction energy effectively to zero (due to the $1/r^6$ dependence of the van der Waals interaction, a 50$\%$ increase in the interatomic distance leads to a reduction in the interaction energy by one order of magnitude). 

When $t_{dark}$ is increased to $5  \,\mathrm{\mu s}$, the effects of thermal motion are clearly visible. The de-excitation resonance at $\Delta=0$  is now more pronounced, whereas those at $\Delta=\Delta_{ex}$ and $\Delta=2\Delta_{ex}$ are substantially reduced. This observation agrees with the fact that for our MOT temperatures the atoms move, on average, $0.6 \,\mathrm{\mu m}$ in $5\,\mathrm{\mu s}$, which leads to a reduction of the interaction energy between excited atoms to around $50\%$ of its initial value.

From the above discussion it is clear that, at least in principle, resonant de-pumping from a Rydberg state can be used as a technique to artificially shorten the lifetime of the Rydberg state, However, the strong dependence of the de-excitation dynamics on Rydberg-Rydberg interactions means that in order for the de-excitation rate to be independent of the spatial distribution of Rydberg atoms (which it must be if it is to implement a true "forced dissipation"), the linewidth of the de-excitation laser has to be much larger than the largest interaction energy to be expected in the system. Since that interaction energy can be on the order of tens of MHz (depending, largely, on the chosen detuning of the excitation lasers), one would have to use a laser with a linewidth of that order of magnitude, necessitating an appropriately high power to compensate for the linewidth. 

\section{The coherent driving regime: signatures of coherent dynamics in non-equilibrium phase transitions}

\subsection{Theoretical results}
The discussion so far focussed entirely on the strongly dissipative limit in which dephasing term in eq. (\ref{eq:master_equation}) represents processes with the fastest timescale. In the future it would certainly be interesting to probe the dynamics and the stationary state of the Rydberg gas in the "coherent" limit, {\em i.e.}, when dephasing can be approximately neglected and radiative decay remains the only significant dissipative process.

The theoretical investigation of this limit is substantially more complicated since the dynamics can no longer be effectively described by a set of classical rate equations. This makes numerical simulations significantly more challenging. In order to get an idea of what to expect in this coherent limit we consider a one-dimensional model for quantum facilitated dynamics with the Hamiltonian \cite{Marcuzzi_2016,buchhold2017}
\begin{eqnarray}
  H=\Omega \sum_k(n_{k-1}+n_{k+1})\sigma_x^k=\sum_k C_k\, \sigma_x^k. \label{eq:q-constraint}
\end{eqnarray}
Here a given spin can change its state, through the operator $\sigma_x^k$, only if at least one of its neighbors is in the excited state, which is probed by the operator $C_k$. While this is an idealised model it has been shown in refs. \cite{Lesanovsky11,Marcuzzi_2017-2,Ostmann_2017} that, indeed, quantum kinetic constraints of this or a similar form are naturally realised within Rydberg gases.

\begin{figure}
\includegraphics[width=14 cm]{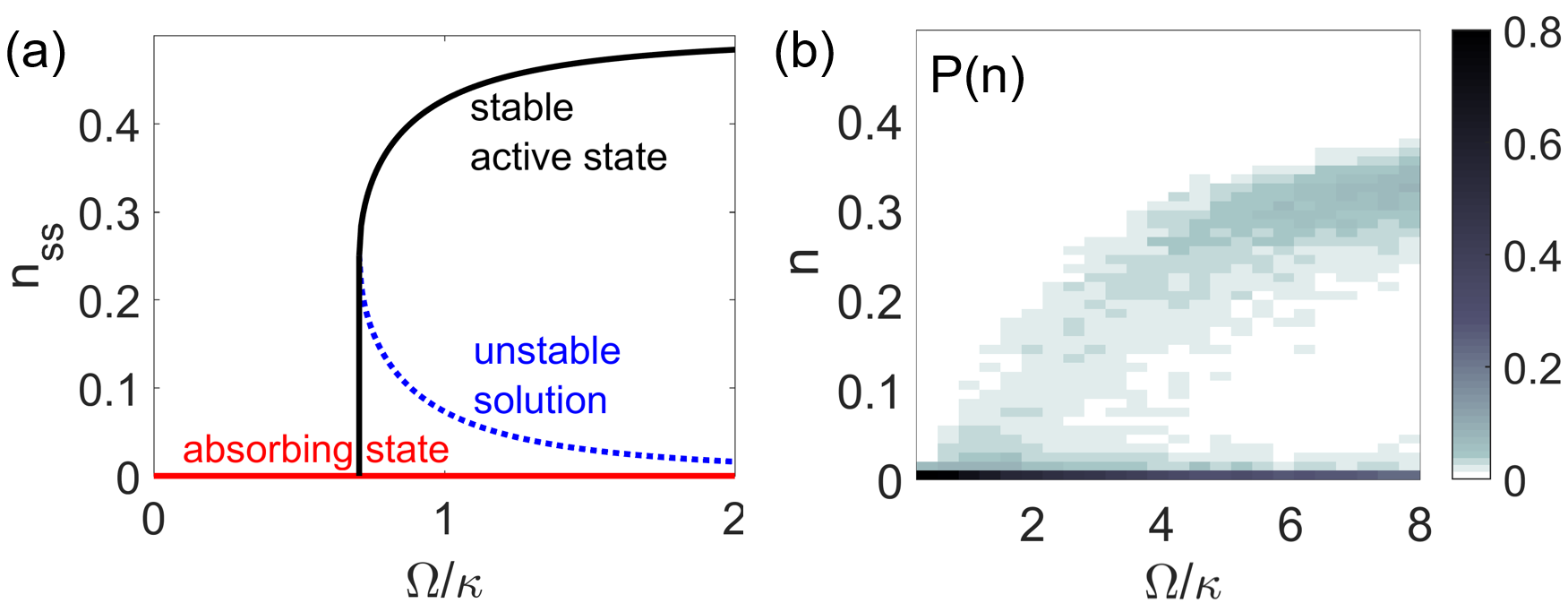}     
\caption{Stationary state phase transition with coherent
facilitation. (a) Stationary density $n_\mathrm{ss}$ with the
meanfield approximation. By construction there is always an
absorbing state solution with $n_\mathrm{ss}=0$. Beyond the
transition point at $\Omega=\kappa/\sqrt{2}$ two further solution
emerge. One is unstable and the second one represents an active
state with finite excitation density $n_\mathrm{ss}$. This result
is suggestive of a first order phase transition. (b) Steady-state
histogram $P(n)$ of the excitation density $n$ for a system
consisting of $12$ spins. The data was obtained via numerically
exact quantum-jump Monte Carlo simulations. For sufficiently large
$\Omega$ a bimodal structure emerges. The two peaks correspond to
the absorbing state and the active state, respectively. This
feature is compatible with the expected first-order character of
the phase transition. For further details the reader is referred to
Refs. \cite{Marcuzzi_2016,buchhold2017}.}\label{fig:QUANTUMTHEORY}
\end{figure}

We proceed by investigating the stationary state that results from a competition between the quantum facilitated dynamics and radiative decay. To this end we substitute Hamiltonian (\ref{eq:q-constraint}) into the Master equation  (\ref{eq:master_equation}), set the dephasing rate $\gamma=0$, and derive a meanfield equation for the stationary excitation density $n_\mathrm{ss}$:
\begin{eqnarray}
  0=n_\mathrm{ss}\left[n_\mathrm{ss}(2n_\mathrm{ss}-1)+\frac{\gamma^2}{16\Omega^2}\right].
\end{eqnarray}
Similar to the classical limit this equation features an absorbing state solution, $n_\mathrm{ss}=0$. Moreover, there are two further solutions, as shown in fig. \ref{fig:QUANTUMTHEORY}a, that emerge beyond a critical driving strength $\Omega\geq\Omega_\mathrm{c}=\kappa/\sqrt{2}$, out of which the stable one is given by
\begin{eqnarray}
  n_\mathrm{ss}=\frac{1}{4}\left[1+\sqrt{1+\frac{\gamma^2}{8\Omega^2}}\right].
\end{eqnarray}
This solution does not smoothly connect to $n_\mathrm{ss}=0$ at the transition point, which indicates the presence a first-order rather than a second-order (continuous) phase transition. This is thus strikingly different to the classical case and numerically exact small scale calculations, conducted via quantum-jump Monte Carlo simulations, appear to be qualitatively consistent with the prediction (see fig. \ref{fig:QUANTUMTHEORY}b).

In refs. \cite{Marcuzzi_2016,buchhold2017} this phase transition, and also the competition between quantum and classical facilitation, was discussed in great detail by employing a field-theoretical approach. This study confirmed the first-order nature in the fully coherent limit, within the approximation employed there. However, it is still a matter of ongoing research to fully characterise the transition. 

\subsection{Towards experimental realizations} To experimentally realize the coherent limit described above, in our experiments it would be necessary to increase the coherence time of the Rydberg excitation or to increase the decay rate, or a combination of both. Increasing the coherence time (currently on the order of a microsecond) is mainly a technical challenge related to the linewidths of the two excitation lasers and the quality of the lock to a Fabry-Perot cavity, which we use to stabilize the wavelengths of the two lasers relative to a reference laser. Recent experiments \cite{Barredo_2016, Bernien_2017} have shown that coherence times of tens of microseconds are achievable, which would take us to within a factor of ten from the spontaneous decay rate of the $70S$ state. Increasing that decay rate is possible using the de-excitation technique described in sec. 3.2, so that, at least in principle, a regime for which $\kappa>\gamma$ should be realizable using state-of-the-art techniques. 

\section{Conclusions} The aim of this review was to show that even in the incoherent, dissipative excitation regime, the many-body dynamics of cold Rydberg gases has intriguing features that make it possible to study classical many-body phenomena. In particular, we demonstrated the existence of kinetic constraints reminiscent of those found in glassy systems, and the emergence of non-equilibrium phase transitions. A profound understanding of those process not only offers opportunities for using cold Rydberg gases as classical many-body simulators, but is also a first step towards developing quantum many-body simulators, which were first proposed by Richard Feynman almost forty years ago \cite{Feynman_1982}. A full characterization of a Rydberg many-body systems containing hundreds of excited atoms (or "spins") in the incoherent regime, with an appropriate validation using the numerical simulations described in this review, will make it possible to make a connection with the coherent regime. To do so, one will have to re-introduce coherent effects (as described in sec. 4.2) in a controlled way by appropriately tuning the coherence and decay times. As some point, numerical simulations will no longer be possible due to the number of spins involved and the increasing importance of the coherences. In that regime, the cold cloud of Rydberg atoms will, for all intents and purposes, be a quantum simulator. The connection with the incoherent regime will then allow one to perfom a "check" or validation in that regime (which can be simulated classically), thus enhancing the reliability of the quantum simulator. First steps in that direction, in which persistent oscillations of magnetic correlations in a chain of interacting and coherently evolving Rydberg atoms were observed, have already been taken \cite{Bernien_2017,Turner_2017}. Also, Rydberg gases can generally be used as a platform for simulating synthetic classical and quantum matter such as quantum glasses \cite{Banerjee2004}, quantum versions of epidemic processes \cite{Perez-Espigares_2017}.  Much work remains to be done before we get there, but it seems to be an achievable - and highly worthwhile - goal to pursue.  

\acknowledgments The work reviewed in this paper was supported by the European Research Council under the European Union's Seventh Framework Programme (FP/2007-2013) / ERC Grant Agreement No. 335266 (ESCQUMA), the EU-FET grant HAIRS 612862 and from the University of Nottingham. Further funding was received through the H2020-FETPROACT-2014 grant No.  640378 (RYSQ) and the EPSRC Grant no.\ EP/M014266/1. IL gratefully acknowledges support through the Royal Society Wolfson Merit Award. OM thanks Cristiano Simonelli for assistance in the preparation of the figures.


\bibliographystyle{varenna}

\bibliography{nuovocimento_bibliography} 

\begin{thebibliography}{10}
\expandafter\ifx\csname url\endcsname\relax\def\url#1{\texttt{#1}}\fi
\expandafter\ifx\csname urlprefix\endcsname\relax\def\urlprefix{URL }\fi

\bibitem{Gallagher_2005}
\NAME{Gallagher T.}, \TITLE{Rydberg {A}toms} (Cambridge University Press) 2005.
\newline\urlprefix\url{https://books.google.it/books?id=8JIpEhHWT-cC}

\bibitem{Cooke_1980}
\NAME{Cooke W.~E. \atque Gallagher T.}, \IN{Phys. Rev. A}{21}{1980}{588}.

\bibitem{Raimond_1981}
\NAME{Raimond J.~M., Vitrant G. \atque Haroche S.}, \IN{J. Phys.
  B}{14}{1981}{L655}.
\newline\urlprefix\url{http://stacks.iop.org/0022-3700/14/i=21/a=003}

\bibitem{saffman_2010}
\NAME{Saffman M., Walker T.~G. \atque M\o{}lmer K.}, \IN{Rev. Mod.
  Phys.}{82}{2010}{2313}.
\newline\urlprefix\url{https://link.aps.org/doi/10.1103/RevModPhys.82.2313}

\bibitem{Isenhower_2010}
\NAME{Isenhower L., Urban E., Zhang X., Gill A., Henage T., Johnson T., Walker
  T. \atque Saffman M.}, \IN{Physical Review Letters}{104}{2010}{}, cited By
  336.
\newline\urlprefix\url{https://www.scopus.com/inward/record.uri?eid=2-s2.0-74949122528
  doi=10.1103}

\bibitem{Barredo_2016}
\NAME{Barredo D., de~L{\'e}s{\'e}leuc S., Lienhard V., Lahaye T. \atque
  Browaeys A.}, \IN{Science}{354}{2016}{1021}.
\newline\urlprefix\url{http://science.sciencemag.org/content/354/6315/1021}

\bibitem{Bernien_2017}
\NAME{Bernien H., Schwartz S., Keesling A., Levine H., Omran A., Pichler H.,
  Choi S., Zibrov A., Endres M., Greiner M., Vuletic V. \atque Lukin M.},
  \IN{Nature}{551}{2017}{579}, cited By 2.
\newline\urlprefix\url{https://www.scopus.com/inward/record.uri?eid=2-s2.0-85036498032
  doi=10.1038}

\bibitem{Comparat_2010}
\NAME{Comparat D. \atque Pillet P.}, \IN{J. Opt. Soc. Am. B}{27}{2010}{A208}.

\bibitem{Simon_2011}
\NAME{Simon J., Bakr W.~S., Ma R., Tai M.~E., Preiss P.~M. \atque Greiner M.},
  \IN{Nature}{472}{2011}{307}.
\newline\urlprefix\url{http://dx.doi.org/10.1038/nature09994}

\bibitem{Bloch_2012}
\NAME{Bloch I., Dalibard J. \atque Nascimb{\`e}ne S.}, \IN{Nat.
  Phys.}{8}{2012}{267}.
\newline\urlprefix\url{http://dx.doi.org/10.1038/nphys2259}

\bibitem{Biroli_2013}
\NAME{Biroli G. \atque Garrahan J.~P.}, \IN{J. Chem. Phys.}{138}{2013}{12A301}.
\newline\urlprefix\url{https://doi.org/10.1063/1.4795539}

\bibitem{Grassberger_1983}
\NAME{Grassberger P.}, \IN{Math. Biosci.}{63}{1983}{157}.

\bibitem{Perez-Espigares_2017}
\NAME{P\'erez-Espigares C., Marcuzzi M., Guti\'errez R. \atque Lesanovsky I.},
  \IN{Phys. Rev. Lett.}{119}{2017}{140401}.
\newline\urlprefix\url{https://link.aps.org/doi/10.1103/PhysRevLett.119.140401}

\bibitem{Lesanovsky_2013}
\NAME{Lesanovsky I. \atque Garrahan J.~P.}, \IN{Phys. Rev.
  Lett.}{111}{2013}{215305}.
\newline\urlprefix\url{https://link.aps.org/doi/10.1103/PhysRevLett.111.215305}

\bibitem{Hinrichsen_2000}
\NAME{Hinrichsen H.}, \IN{Adv. Phys.}{49}{2000}{815}.
\newline\urlprefix\url{http://dx.doi.org/10.1080/00018730050198152}

\bibitem{Hinrichsen_2006}
\NAME{Hinrichsen H.}, \IN{Physica A}{369}{2006}{1}.
\newline\urlprefix\url{http://dx.doi.org/10.1016/j.physa.2006.04.007}

\bibitem{Valado_2015}
\NAME{Valado M.~M., Hoogerland M.~D., Simonelli C., Arimondo E., Ciampini D.
  \atque Morsch O.}, \IN{J. Phys. Conf. Ser.}{605}{2015}{012038}.

\bibitem{Simonelliphd_2018}
\NAME{Simonelli C.}, Ph.D. thesis, University of Pisa (2018).

\bibitem{Viteau_2011}
\NAME{Viteau M., Radogostowicz J., Bason M.~G., Malossi N., Ciampini D., Morsch
  O. \atque Arimondo E.}, \IN{Opt. Express}{19}{2011}{6007}.
\newline\urlprefix\url{http://www.opticsexpress.org/abstract.cfm?URI=oe-19-7-6007}

\bibitem{Pfau_2012}
\NAME{L{\"o}w R., Weimer H., Nipper J., Balewski J., Butscher B., B{\"u}chler
  H.~P. \atque Pfau T.}, \IN{J. Phys. B}{45}{2012}{113001}.

\bibitem{Malossi_2014}
\NAME{Malossi N., Valado M.~M., Scotto S., Huillery P., Pillet P., Ciampini D.,
  Arimondo E. \atque Morsch O.}, \IN{Phys. Rev. Lett.}{113}{2014}{023006}.
\newline\urlprefix\url{https://link.aps.org/doi/10.1103/PhysRevLett.113.023006}

\bibitem{Cai13}
\NAME{Cai Z. \atque Barthel T.}, \IN{Phys. Rev. Lett.}{111}{2013}{150403}.
\newline\urlprefix\url{https://link.aps.org/doi/10.1103/PhysRevLett.111.150403}

\bibitem{Marcuzzi14}
\NAME{Marcuzzi M., Schick J., Olmos B. \atque Lesanovsky I.}, \IN{J. Phys. A:
  Math. Theor.}{47}{2014}{482001}.

\bibitem{Degenfeld14}
\NAME{Degenfeld-Schonburg P. \atque Hartmann M.~J.}, \IN{Phys. Rev.
  B}{89}{2014}{245108}.
\newline\urlprefix\url{http://link.aps.org/doi/10.1103/PhysRevB.89.245108}

\bibitem{Gutierrez15}
\NAME{Guti\'errez R., Garrahan J.~P. \atque Lesanovsky I.}, \IN{Phys. Rev.
  E}{92}{2015}{062144}.
\newline\urlprefix\url{https://link.aps.org/doi/10.1103/PhysRevE.92.062144}

\bibitem{Lesanovsky_2014}
\NAME{Lesanovsky I. \atque Garrahan J.~P.}, \IN{Phys. Rev.
  A}{90}{2014}{011603(R)}.
\newline\urlprefix\url{http://dx.doi.org/10.1103/PhysRevA.90.011603}

\bibitem{Valado_2016}
\NAME{Valado M.~M., Simonelli C., Hoogerland M.~D., Lesanovsky I., Garrahan
  J.~P., Arimondo E., Ciampini D. \atque Morsch O.}, \IN{Phys. Rev.
  A}{93}{2016}{040701(R)}.
\newline\urlprefix\url{http://dx.doi.org/10.1103/PhysRevA.93.040701}

\bibitem{Urvoy_2015}
\NAME{Urvoy A., Ripka F., Lesanovsky I., Booth D., Shaffer J.~P., Pfau T.
  \atque L\"ow R.}, \IN{Phys. Rev. Lett.}{114}{2015}{203002}.
\newline\urlprefix\url{http://link.aps.org/doi/10.1103/PhysRevLett.114.203002}

\bibitem{Heidemann_2007}
\NAME{Heidemann R.}, \IN{Phys. Rev. Lett.}{99}{2007}{}.

\bibitem{Dudin_2012}
\NAME{Dudin Y.~O., Li L., Bariani F. \atque Kuzmich A.}, \IN{Nat.
  Phys.}{8}{2012}{790}.
\newline\urlprefix\url{http://dx.doi.org/10.1038/nphys2413}

\bibitem{Schauss_2012}
\NAME{Schau{\ss} P., Cheneau M., Endres M., Fukuhara T., Hild S., Omran A.,
  Pohl T., Gross C., Kuhr S. \atque Bloch I.}, \IN{Nature}{491}{2012}{87}.
\newline\urlprefix\url{http://dx.doi.org/10.1038/nature11596}

\bibitem{Viteau_2012}
\NAME{Viteau M., Huillery P., Bason M.~G., Malossi N., Ciampini D., Morsch O.,
  Arimondo E., Comparat D. \atque Pillet P.}, \IN{Phys. Rev.
  Lett.}{109}{2012}{053002}.

\bibitem{Gaetan_2009}
\NAME{Ga{\"e}tan A., Miroshnychenko Y., Wilk T., Chotia A., Viteau M., Comparat
  D., Pillet P., Browaeys A. \atque Grangier P.}, \IN{Nat.
  Phys.}{5}{2009}{115}.

\bibitem{Urban_2009}
\NAME{Urban E., Johnson T.~A., Henage T., Isenhower L., Yavuz D.~D., Walker
  T.~G. \atque Saffman M.}, \IN{Nat. Phys.}{5}{2009}{110}.

\bibitem{Garttner_2013}
\NAME{G\"arttner M., Heeg K.~P., Gasenzer T. \atque Evers J.}, \IN{Phys. Rev.
  A}{88}{2013}{043410}.
\newline\urlprefix\url{https://link.aps.org/doi/10.1103/PhysRevA.88.043410}

\bibitem{Schempp_2014}
\NAME{Schempp H., G\"unter G., Robert-de Saint-Vincent M., Hofmann C.~S.,
  Breyel D., Komnik A., Sch\"onleber D.~W., G\"arttner M., Evers J., Whitlock
  S. \atque Weidem\"uller M.}, \IN{Phys. Rev. Lett.}{112}{2014}{013002}.
\newline\urlprefix\url{https://link.aps.org/doi/10.1103/PhysRevLett.112.013002}

\bibitem{Simonelli_2016}
\NAME{Simonelli C., Valado M.~M., Masella G., Asteria L., Arimondo E., Ciampini
  D. \atque Morsch O.}, \IN{J. Phys. B}{49}{2016}{154002}.
\newline\urlprefix\url{http://stacks.iop.org/0953-4075/49/i=15/a=154002}

\bibitem{Marcuzzi_2015}
\NAME{Marcuzzi M., Levi E., Li W., Garrahan J.~P., Olmos B. \atque Lesanovsky
  I.}, \IN{New J. Phys.}{17}{2015}{072003}.
\newline\urlprefix\url{http://dx.doi.org/10.1088/1367-2630/17/7/072003}

\bibitem{Gutierrez_2017}
\NAME{Guti\'errez R., Simonelli C., Archimi M., Castellucci F., Arimondo E.,
  Ciampini D., Marcuzzi M., Lesanovsky I. \atque Morsch O.}, \IN{Phys. Rev.
  A}{96}{2017}{041602(R)}.
\newline\urlprefix\url{https://link.aps.org/doi/10.1103/PhysRevA.96.041602}

\bibitem{Everest_2016}
\NAME{Everest B., Marcuzzi M. \atque Lesanovsky I.}, \IN{Phys. Rev.
  A}{93}{2016}{023409}.
\newline\urlprefix\url{https://link.aps.org/doi/10.1103/PhysRevA.93.023409}

\bibitem{Simonelli_2017}
\NAME{Simonelli C., Archimi M., Asteria L., Capecchi D., Masella G., Arimondo
  E., Ciampini D. \atque Morsch O.}, \IN{Phys. Rev. A}{96}{2017}{043411}.
\newline\urlprefix\url{https://link.aps.org/doi/10.1103/PhysRevA.96.043411}

\bibitem{Faoro_2016}
\NAME{Faoro R., Simonelli C., Archimi M., Masella G., Valado M.~M., Arimondo
  E., Mannella R., Ciampini D. \atque Morsch O.}, \IN{Phys. Rev.
  A}{93}{2016}{030701(R)}.
\newline\urlprefix\url{http://dx.doi.org/10.1103/PhysRevA.93.030701}

\bibitem{Shore_2006}
\NAME{Shore B.~W. \atque Vitanov N.~V.}, \IN{Contemp. Phys.}{47}{2006}{341}.

\bibitem{Thaicharoen_2015}
\NAME{Thaicharoen N., Schwarzkopf A. \atque Raithel G.}, \IN{Phys. Rev.
  A}{92}{2015}{040701(R)}.
\newline\urlprefix\url{https://link.aps.org/doi/10.1103/PhysRevA.92.040701}

\bibitem{Teixeira_2015}
\NAME{Teixeira R.~C., Hermann-Avigliano C., Nguyen T.~L., Cantat-Moltrecht T.,
  Raimond J.~M., Haroche S., Gleyzes S. \atque Brune M.}, \IN{Phys. Rev.
  Lett.}{115}{2015}{013001}.
\newline\urlprefix\url{https://link.aps.org/doi/10.1103/PhysRevLett.115.013001}

\bibitem{Marcuzzi_2016}
\NAME{Marcuzzi M., Buchhold M., Diehl S. \atque Lesanovsky I.}, \IN{Phys. Rev.
  Lett.}{116}{2016}{245701}.
\newline\urlprefix\url{http://link.aps.org/doi/10.1103/PhysRevLett.116.245701}

\bibitem{buchhold2017}
\NAME{Buchhold M., Everest B., Marcuzzi M., Lesanovsky I. \atque Diehl S.},
  \IN{Phys. Rev. B}{95}{2017}{014308}.

\bibitem{Lesanovsky11}
\NAME{Lesanovsky I.}, \IN{Phys. Rev. Lett.}{106}{2011}{025301}.
\newline\urlprefix\url{https://link.aps.org/doi/10.1103/PhysRevLett.106.025301}

\bibitem{Marcuzzi_2017-2}
\NAME{Marcuzzi M., Min\'a\ifmmode~\check{r}\else \v{r}\fi{} J. c.~v., Barredo
  D., de~L\'es\'eleuc S., Labuhn H., Lahaye T., Browaeys A., Levi E. \atque
  Lesanovsky I.}, \IN{Phys. Rev. Lett.}{118}{2017}{063606}.
\newline\urlprefix\url{https://link.aps.org/doi/10.1103/PhysRevLett.118.063606}

\bibitem{Ostmann_2017}
\NAME{Ostmann M., Minář J., Marcuzzi M., Levi E. \atque Lesanovsky I.},
  \IN{New J. Phys.}{}{2017}{123015}.

\bibitem{Feynman_1982}
\NAME{Feynman R.~P.}, \IN{Int. J. Theor. Phys.}{21}{1982}{467}.
\newline\urlprefix\url{https://doi.org/10.1007/BF02650179}

\bibitem{Turner_2017}
\NAME{{Turner} C.~J., {Michailidis} A.~A., {Abanin} D.~A., {Serbyn} M. \atque
  {Papic} Z.}, \TITLE{{Quantum many-body scars}} (Nov. 2017).

\bibitem{Banerjee2004}
\NAME{Banerjee V. \atque Dattagupta S.}, \IN{Phase Transitions}{77}{2004}{525},
  cited By 3.
\newline\urlprefix\url{https://www.scopus.com/inward/record.uri?eid=2-s2.0-11144349059
  doi=10.1080}

\end{thebibliography}

\end{document}